\documentclass[useAMS,usenatbib]{mn2e}

\usepackage{natbib, aas_macros}
\usepackage{version}
\usepackage{ulem}
\usepackage{graphicx}
\usepackage{wrapfig}
\graphicspath{{./figs/}}
\usepackage{color}
\usepackage{amsmath}
\usepackage{amssymb}

\newcommand{\Msun}{{\rm  M_{\odot}}}

\newcommand{\Zsun}{Z_{\odot}}

\newcommand{\Lsun}{L_{\odot}}

\newcommand{\lya}{\rm {Ly{\alpha}}}

\newcommand{\Mh}{M_{\rm h}}

\newcommand{\un}[2]{#1_{\rm #2}}

\newcommand{\artart}{${\rm ART^{2}}$ }

\title[Radiative properties of the first galaxies] 
{
Radiative properties of the first galaxies: rapid transition between blue and red
}
\author[Arata et al.]
{Shohei Arata$^{1}$\thanks{E-mail: arata@astro-osaka.jp},
Hidenobu Yajima$^{2}$,
Kentaro Nagamine$^{1,3,4}$,
Yuexing Li$^{5,6}$\and
and Sadegh Khochfar$^{7}$
\\
$^{1}$ Department of Earth and Space Science, Graduate School of Science, Osaka University, Toyonaka, Osaka 560-0043, Japan\\
$^{2}$ Center of Computational Sciences University of Tsukuba, Ibaraki 305-8577, Japan\\
$^{3}$ Department of Physics \& Astronomy, University of Nevada, Las Vegas, 4505 S. Maryland Pkwy, Las Vegas, NV 89154-4002, USA \\
$^{4}$ Kavli IPMU, The University of Tokyo, 5-1-5 Kashiwanoha, Kashiwa, Chiba, 277-8583, Japan \\
$^{5}$ Department of Astronomy \& Astrophysics, The Pennsylvania State University, 525 Davey Lab, University Park, PA 16802, USA \\
$^{6}$ Institute for Gravitation and the Cosmos, The Pennsylvania State University, University Park, PA 16802, USA \\
$^{7}$ SUPA, Institute for Astronomy, University of Edinburgh, Royal Observatory, Edinburgh, EH9 3HJ, UK
}
 
\begin{document}

\date{Accepted ?; Received ??; in original form ???}

\pagerange{\pageref{firstpage}--\pageref{lastpage}} \pubyear{2008}

\maketitle

\label{firstpage}

%
%
\begin{abstract}
Recent observations have successfully detected UV-bright and infrared-bright galaxies in the epoch of reionization. 
However, the origin of their radiative properties has not been understood yet. 
Combining cosmological hydrodynamic simulations and radiative transfer calculations, we present predictions of multi-wavelength radiative properties of the first galaxies at $z\sim 6-15$.
Using zoom-in initial conditions, we investigate three massive galaxies and their satellites in different environment and halo masses: $\un{M}{h} = 2.4\times 10^{10}\Msun$ (Halo-10), $1.6\times 10^{11}\Msun$ (Halo-11) and $0.7\times 10^{12}\Msun$ (Halo-12) at $z=6$.
We find that most of gas and dust are ejected  from star-forming regions by supernova feedback, which allows UV photons to escape.
We show that the peak of the spectral energy distribution (SED) rapidly changes between UV and infrared wavelengths on a time-scale of $\sim$\,100\,Myrs due to intermittent star formation and feedback, and the escape fraction of UV photons fluctuates in the range of $0.2-0.8$ at $z<10$ with a time-averaged value of 0.3. 
When dusty gas covers the star-forming regions, the galaxies become bright in the observed-frame sub-millimeter wavelengths. 
We predict the detectability of high-$z$ galaxies with the Atacama Large Millimeter Array (ALMA). 
For a sensitivity limit of $0.1\,{\rm mJy}$ at $850\,{\rm \mu m}$, the detection probability of galaxies in halos $\Mh \gtrsim 10^{11}\,\Msun$ at $z\lesssim 7$ exceeds fifty per cent. 
We argue that supernova feedback can produce the observed diversity of SEDs for high-$z$ galaxies.
\end{abstract}

%
%


%
%

\section{Introduction}
\label{intro}

Understanding galaxy evolution is one of the main goals of modern astronomy and astrophysics.
Recent observations have detected many galaxies in the early Universe, the so-called {\it first galaxies}, in the optical and near-infrared wavelengths \citep[e.g.][]{Ouchi09, Ono12, Shibuya12, McLure13, Finkelstein13, Finkelstein15, Bouwens15, Oesch16, Ono17, Bowler18}, or in the sub-millimeter wavelengths \citep{Riechers13, Watson15, Inoue16, Marrone18, Hashimoto18b}. 

In observational studies of high-$z$ galaxies, a large number of star-forming galaxies are usually color-selected first as Lyman-break galaxies \citep[LBGs; e.g.,][]{Steidel96,Steidel03,Shapley03,Bouwens15, Oesch16, Bowler18}, 
and then later their redshifts are confirmed by spectroscopy. 
Many high-$z$ galaxies at $z\gtrsim 6$ are also detected by the Ly$\alpha$ line \citep[LAEs; e.g.,][]{Ouchi03,Hu04,Dawson04,Shimasaku06,Kashikawa06,Gronwall07,Ouchi09,Ouchi10}.  
In addition, ALMA observations are beginning to provide valuable information on high-$z$ galaxies via detections of dust continuum, metal lines and molecular lines \citep[e.g.,][]{Inoue16, Hashimoto18b}, so-called sub-millimeter galaxies \citep[SMGs;][]{Marrone18}.
For example, \citet{Riechers13} showed that the dusty massive starburst galaxies are already formed as early as at $z\sim 6$.  
These different varieties of high-$z$ population and radiative properties probably reflect different physical conditions in the first galaxies.  Therefore studying the radiative output from the first galaxies is one of the primary ways to understand galaxy formation and evolution. 

The formation mechanism of the first galaxies has been studied by numerical simulations \citep{Greif08,Maio11,Safranek-Shrader12,Wise12a,Johnson13,Hopkins14,Kimm14,Yajima15b,Paardekooper15,Yajima17,Ricotti16,Trebitsch17,Kim17}. 
These studies reveal that stellar and supernova (SN) feedback are crucial for galaxy evolution.
In particular, \cite{Yajima17} studied galaxy evolution in massive halos ($\Mh \gtrsim 10^{11}~\Msun$) at $z\gtrsim 6$ using cosmological hydrodynamic simulations, and found that the star formation occurs intermittently due to SN feedback and gas accretion \citep[see also,][]{Kimm14}.
However, the radiative properties associated with the bursty star-formation history are not fully understood yet. 
In this paper, we focus on the radiative output of the first galaxies by carrying out radiative transfer calculations and post-processing the zoom-in hydrodynamic simulations.  

In order to study the radiative properties, we need to know the spatial distribution of dust and stars.
As galaxies evolve, dust and metals are produced via Type-II SNe.\footnote{AGB stars also produce dust. However, the evolution of low-mass stars toward the AGB phase takes longer than $1$\,Gyr. Thus, SNe dominate the dust enrichment of first galaxies, which occurs at the end of the lifetimes of massive stars ($\lesssim 10$\,Myr). The pair-instability SNe are also important in the early universe \citep[e.g.][]{Nozawa07}, but for evolved galaxies Type-II SNe dominate the dust production. For example, \cite{Maiolino04} showed that the extinction curves of $z\sim 6$ quasars agree with the Type-II SN model.
}
When the interstellar dust absorbs the stellar UV radiation and re-emits the energy in the infrared, the galaxy can be bright in the sub-mm wavelengths due to the thermal emission from dust.
In contrast, if the dust absorption is ineffective, star-forming galaxies are bright in the rest-frame UV wavelengths and become targets for Hubble Space Telescope (HST), Subaru and Keck telescopes. 

Dust attenuation sensitively  depends on the dust abundance, size distribution and spatial distribution.
However, the information on the dust properties from observations is limited, and theoretical studies can give important insight.  
For example, \citet{Todini01} theoretically studied the dust size and composition by taking into account the dust growth/destruction processes in supernova shock \citep[see also,][]{Schneider06,Nozawa07,Chiaki13,Chiaki15}.
\citet{Asano13} evaluated the dust properties using phenomenological models which considered dust formation and destruction processes in combination with star formation histories in galaxies. 
\citet{Yajima14a} carried out radiative transfer calculations in cosmological simulations, and estimated the typical dust size by comparing the colors of simulated galaxies and observations. 
\cite{Cullen17} examined the dust attenuation law by matching the UV-LF and UV-beta slope of simulated galaxies, and showed that the Calzetti law is actually the best description rather than the SMC law \citep[see also][]{Cullen18}.
More recently, \cite{Narayanan18b} showed that the dust extinction curve varied due to the inhomogeneous spatial distribution of dust for star-forming galaxies at $z=0-6$ in cosmological simulations \citep[see also,][]{Hou17}. 
In addition, \cite{Behrens18} studied the temperature and amount of dust in the observed SMG at $z=8.38$ \citep{Laporte17} using  cosmological simulations.
In the present work, we study the sub-mm fluxes from more massive galaxies with bursty star formation histories and the impact of stellar feedback on the radiative properties of first galaxies at $z\gtrsim 6$ using cosmological zoom-in hydrodynamic simulations. 

In addition to the SED of galaxies, the size of galaxies at UV wavelengths is likely to change as galaxies evolve. 
Galaxy sizes have a large impact on the observational estimation of the faint-end slope of UV luminosity function, because of surface brightness dimming and observational limits \citep{Grazian11}.   
As we discuss in this paper, the morphology and the extension of galactic disk is significantly dependent on the feedback strength at high-$z$, and if the extended low-mass galaxies are missed in the current Hubble Frontier Fields observation, the cosmic SFRD could be significantly underestimated at $z\gtrsim 6$ \citep[e.g.][]{Jaacks12a}.  The low-mass galaxies studied in this paper are mainly satellites of massive galaxies at $z\gtrsim 6$, and separate zoom simulations of low-density regions will have to be performed in the future to examine the low-mass galaxies in the field regions. 
In any case, we investigate the galaxy sizes as well as SEDs by radiative transfer calculations, and find an interesting transition between UV to sub-mm bright phase due to intermittent starburst and its feedback. This fluctuation may affect the estimate of luminosity function as we will discuss in Sec.~\ref{sec:flux}, and we quantify the duty cycle of such fluctuations. 

Our paper is organized as follows.
We describe the methodologies of cosmological hydrodynamic simulations and radiative transfer calculations in Section~\ref{sec:method}.
In Section~\ref{sec:result}, we present our results.
In Sections~\ref{sec:Halo-11} and \ref{sec:flux}, we focus on the fluctuation of UV escape fraction and sub-mm flux of dust re-emission, respectively.
In Section~\ref{sec:duty}, we present the detectability of first galaxies by ALMA. 
In Sections~\ref{sec:dust} and \ref{sec:size}, we show the dust temperature and half-light radius of galaxies at UV wavelength. 
In addition, we discuss the dependence of our results on the the models of star formation and feedback in Section~\ref{sec:discussion}.
Finally we summarize our main conclusions in Section~\ref{sec:summary}.


%
%
\section{Method}
\label{sec:method}

\subsection{Cosmological hydrodynamic simulations}
\label{sec:hydro}

\begin{table*}
  \centering
  \begin{tabular}{ccccccc}
    \hline
    Halo ID  & $\un{M}{h}~[h^{-1}~\Msun]$  &  $\un{m}{DM}~[h^{-1}~\Msun]$  &  $\un{m}{gas}~[h^{-1}~\Msun]$  &  $\un{\epsilon}{min}~[h^{-1}~{\rm pc}]$  &  SNe feedback  &  $A$\\
    \hline \hline
    Halo-10  & $2.4\times 10^{10}$  & $6.6\times 10^{4}$ & $1.2\times 10^{4}$ & $200$ & ON & $2.5\times 10^{-3}$ \\
    Halo-11  & $1.6\times 10^{11}$  & $6.6\times 10^{4}$ & $1.2\times 10^{4}$ & $200$ & ON & $2.5\times 10^{-3}$ \\
    Halo-12  & $7.5\times 10^{11}$  & $1.1\times 10^{6}$ & $1.8\times 10^{5}$ & $200$ & ON & $2.5\times 10^{-3}$ \\
    Halo-11-lowSF  & $1.6\times 10^{11}$  & $6.6\times 10^{4}$ & $1.2\times 10^{4}$ & $200$ & ON & $2.5\times 10^{-4}$ \\
    Halo-11-noSN  & $1.6\times 10^{11}$  & $6.6\times 10^{4}$ & $1.2\times 10^{4}$ & $200$ & OFF & $2.5\times 10^{-3}$ \\
    \hline
  \end{tabular}
\caption{Parameters of our zoom-in cosmological hydrodynamic simulations: (1) $\un{M}{h}$ is the halo mass at $z = 6$. (2) $\un{m}{DM}$ is the mass of a dark matter particle. (3) $\un{m}{gas}$ is the initial mass of a gas particle. (4) $\un{\epsilon}{min}$ is the gravitational softening length in comoving units. (5) $A$ is the amplitude factor for the star formation model based on the Kennicutt-Schmidt law \citep{Schaye08}.
The Halo-11-lowSF run has a lower star formation amplitude factor. The Halo-11-noSN run has no SN feedback. 
 }
\label{table:setup}
\end{table*}

We use the same zoom-in cosmological hydrodynamic simulation as in \citet[][hereafter Y17]{Yajima17}, who focused on three haloes in different environment with halo masses $2.4\times 10^{10}~h^{-1}\Msun$ (Halo-10), $1.6\times 10^{11}~h^{-1}\Msun$ (Halo-11) and $7.5\times 10^{11}~h^{-1}\Msun$ (Halo-12) at $z=6$. 
The volumes of the entire simulation boxes are $(20~h^{-1}{\rm Mpc})^3$ for Halo-10 and Halo-11, and $(100~h^{-1}{\rm Mpc})^3$ for Halo-12 in comoving units. 
In order to increase the resolution, Y17 used the zoom-in initial conditions for the regions of $(1-2~h^{-1}{\rm Mpc})^3$ for Halo-10 and Halo-11 and $(6.6~h^{-1}{\rm Mpc})^3$ for Halo-12, which achieved the resolution of  $\sim 10-30~\rm pc$ in physical units at $z \sim 10$. 
With this resolution, the ISM structure within galaxies are resolved down to the scales of just above molecular clouds, and the results of our radiative transfer calculation is more reliable than those using cosmological simulations with resolution coarser than 100\,pc as we describe in the next subsection. 
Other parameters are shown in Table~\ref{table:setup}.  

\begin{figure*}
\begin{center}
\includegraphics[width=2\columnwidth]{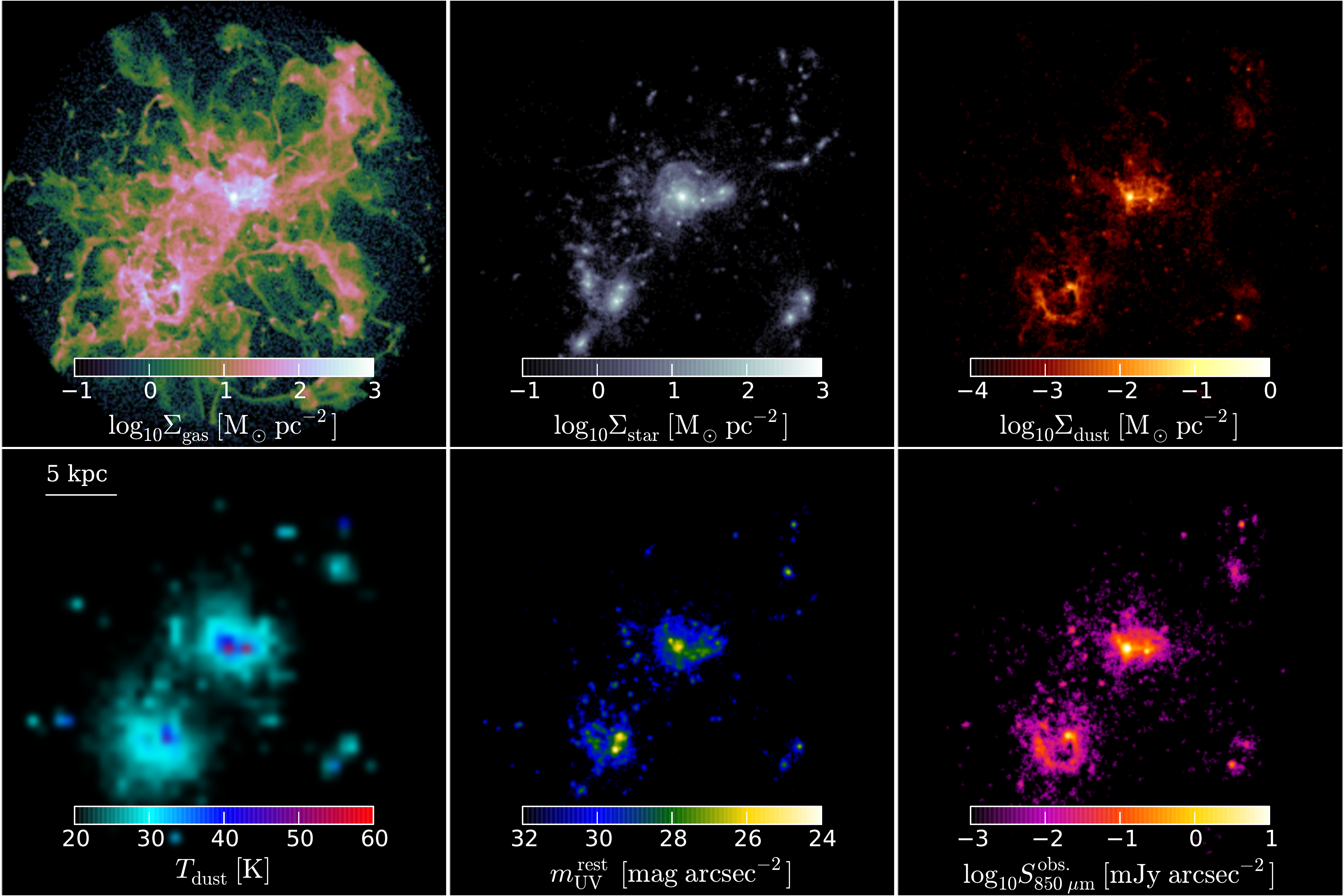}
\caption{
Maps of the main galaxy in Halo-11 run at $z\sim 6$. 
{\it Top panels:} Column density of gas (left), stars (middle) and dust (right).
{\it Bottom panels:} Dust temperature (left), surface brightness of the rest-frame UV continuum (middle) and the observed-frame sub-mm continuum (right). The pixel size is $\sim 0.02~{\rm arcsec}$.
The spatial scale of 5\,kpc (physical) is displayed in the bottom left panel.
}
\label{fig:maps}
\end{center}
\end{figure*}

The upper panels of Figure~\ref{fig:maps} show the distribution of gas, stars and dust in Halo-11 at $z\sim 6$.
We calculate the dust amount from gas metallicity (see Section~\ref{sec:rt}).
In our simulations, we use the star formation model based on Kennicutt-Schmidt law \citep{Schaye08}
and the stochastic thermal feedback model for supernova (SN) feedback \citep{Dalla12}.
For Halo-10, Halo-11, and Halo-12 runs, we use the same parameter set for star formation and feedback models. 
Y17 also compared the evolution of Halo-11 in the case of lower star-formation efficiency (Halo-11-lowSF) and the case without SN feedback (Halo-11-noSN).
We examine the radiative properties of first galaxies in these different haloes in Section~\ref{sec:discussion}.

%
%
\subsection{Radiative transfer}
\label{sec:rt}

We use the radiative transfer code, All-wavelength Radiative Transfer with Adaptive Refinement Tree \citep[$\rm ART^2$:][]{Li08, Yajima12}. 
Here we briefly explain the features of the code.

\artart is based on a Monte Carlo technique.
It tracks the propagation of photon packets emitted from star particles, and computes the emergent SED of a galaxy.
The intrinsic SED of each star particle before dust extinction is calculated by {\sc Starburst99} with the information of stellar age and metallicity.
We assume the Chabrier IMF with the mass range from $0.1$ to $100~M_{\odot}$ \citep{Chabrier03}.    For each galaxy, we cast $10^6$ photon packets, which is sufficient to achieve a reasonable convergence, based on the comparison with the case of $10^5$ photon packets. 
\artart uses an adaptive refinement grid structure. 
We initially construct $4^3$ base-grid over the zoom region which covers twice the size of virial radius. 
If the number of SPH particles in a cell is greater than 16, the cell is further refined by $2^3$ grid.
We set the maximum refinement level to 12 which achieves the resolution of $2.7~h^{-1}{\rm pc}$ for Halo-11 at $z=6$. 

Even in the current resolution, it is difficult to resolve the multi-phase ISM accurately. 
Thermal instability is one of the main processes to form  multi-phase ISM \citep[e.g.][]{Field65,MO77}, and it can occur even in the high-$z$ galaxies with low metallicities \citep[e.g.][]{IO15}.
\citet{Arata18} found that the spatial resolution of  $\lesssim 0.01~{\rm pc}$ was required to follow the formation of dense cold clumps, which is still difficult to achieve even with the state-of-the-art cosmological zoom simulations.
Therefore, \artart calculates the radiative transfer with a sub-grid multi-phase ISM model based on \citet{SH03}, which computes cold dense cloud mass fraction within the hot phase assuming an equilibrium. 

We assume that the dust-to-gas mass ratio $\mathcal{D}$ in the cold clouds is proportional to the local metallicity as in the local galaxies, $\mathcal{D}=8\times10^{-3}~(Z/\Zsun)$ \citep{Draine07}. 
On the other hand, $\mathcal{D}$ in the hot gas is scaled by the hydrogen neutral fraction considering that the dust is efficiently destroyed in the ionized regions. 
When the ionized gas recombines, it is split into two phases again. For example, in a cold cloud with $n_{\rm H}=10^{3}~{\rm cm^{-3}}$ and $T=50~{\rm K}$, the time-scale of dust growth via accretion is $\sim 10(Z/0.1\Zsun)~{\rm Myr}$ \citep{Hirashita11}, which roughly corresponds to the time interval between our snapshots. Thus the dust reformation proceeds fast enough, and the total dust mass is dominated by the dust in the cold clouds.

We adopt the dust size distribution of \cite{Todini01} for solar metallicity and $M=22~\Msun$ SN model.
The dust opacity curve is calculated by combining the size distribution with the cross section of \cite{Weingartner01}.  

To obtain panchromatic SEDs, 
\artart treats the transfer of photons with various frequencies. 
First, it tracks the propagation of ionizing photons.
Then, using the obtained ionized structure, it calculates the transfer of UV continuum and dust absorption/re-emission.
When the dust absorbs photon packets within a time interval $\Delta t$, the dust temperature is updated under the assumption of radiative equilibrium: 
\begin{equation}
E_{\rm abs} = 4\pi\, \Delta t\, \kappa_{\rm P}(T_{\rm d})\, B(T_{\rm d})\, m_{\rm d},
\end{equation}
where $\kappa_{\rm P}=\int \kappa_{\nu} B_{\nu}d \nu/ \int B_{\nu}d \nu$ is the Planck mean opacity, and $B_{\nu}$ is the black-body radiation. 
After the determination of dust temperature, we calculate the radiative transfer of infrared photons from dust.


%
%
\section{Result} 
\label{sec:result}


\subsection{Projected Images}
\label{sec:Halo-11}
Figure~\ref{fig:maps} shows various projection plots of Halo-11 at $z=6$ including UV and sub-mm surface brightness. 
The total stellar mass, gas mass, and dust mass in this halo are $M_\star = 2.4\times 10^{9}\,h^{-1}\Msun$, $M_{\rm gas}=2.5\times 10^{10}\,h^{-1}\Msun$, and $M_{\rm dust}=2\times 10^{7}\,h^{-1}\Msun$. 

The top left panel shows the total gas column density, which is extended over about $20$\,kpc (physical) with complex filamentary structure.  The gas distribution is disturbed by multiple galaxy mergers and high gas inflow rate, and its motion is turbulent.  The green, pink, and white regions correspond to hydrogen column densities of $N_{\rm H} \sim 10^{20}, 10^{21}, 10^{22}\,{\rm cm}^{-2}$, respectively. The pink and white regions would certainly correspond to damped Ly$\alpha$ systems (DLAs), which will produce a deep absorption trough if we were to have a bright quasar behind this system as a background source. 

The top middle panel shows a massive stellar system in the center of this halo with $M_{\star} \sim 10^{9} h^{-1}\Msun$, and there is another relatively large satellite galaxy in the lower left side of the panel with $M_{\star} \sim 3\times 10^{8} h^{-1}\Msun$ which will soon merge with the central system, where we identify satellite galaxies with {\sc subfind} algorithm \citep{Springel01}.  The middle bottom panel shows the UV surface brightness, which traces the young stellar population.
Here, the brightest white pixels have $m_{\rm AB}\sim 24.1~{\rm mag~arcsec^{-2}}$, corresponding to the local projected star formation rate of about $49\,\Msun\,{\rm yr^{-1}\,arcsec}^{-2}$.  

The top right panel shows the projected dust mass distribution, which largely traces the stellar distribution, but with interesting offsets on small scales.  
The observed sub-mm surface brightness reflects the underlying dust distribution, and 850\,$\mu$m surface brightness is shown in the bottom right panel. 
Here, the brightest pixel has $0.84\,{\rm mJy\,arcsec^{-2}}$, and most of the UV radiation are absorbed by dust in such a region. 
The brightest pixel of observed sub-mm band is in the main galaxy, meanwhile that of rest-UV band is in the satellite galaxy on the lower left side of the panel, resulting in a large offset of $\sim 1.4\,$arcsec between them, which corresponds to a physical separation of $8$\,kpc at $z=6$.
Such an offset has been discovered in the observed galaxies at $z\sim 7$ \citep{Bowler18,Hashimoto18b}.
This suggests that some star-forming regions are obscured by dust,  while others are optically thin to the UV radiation.

In the right-column panels, there is also an interesting  bubble-like structure around the satellite galaxy on the lower left side of the panel, which presumably was caused by the SN feedback from nearby starburst.  We will study the dynamics of these SN bubbles in the future furthermore.

\subsection{SFR and Escape Fraction}

\begin{figure}
\begin{center}
\includegraphics[width=\columnwidth]{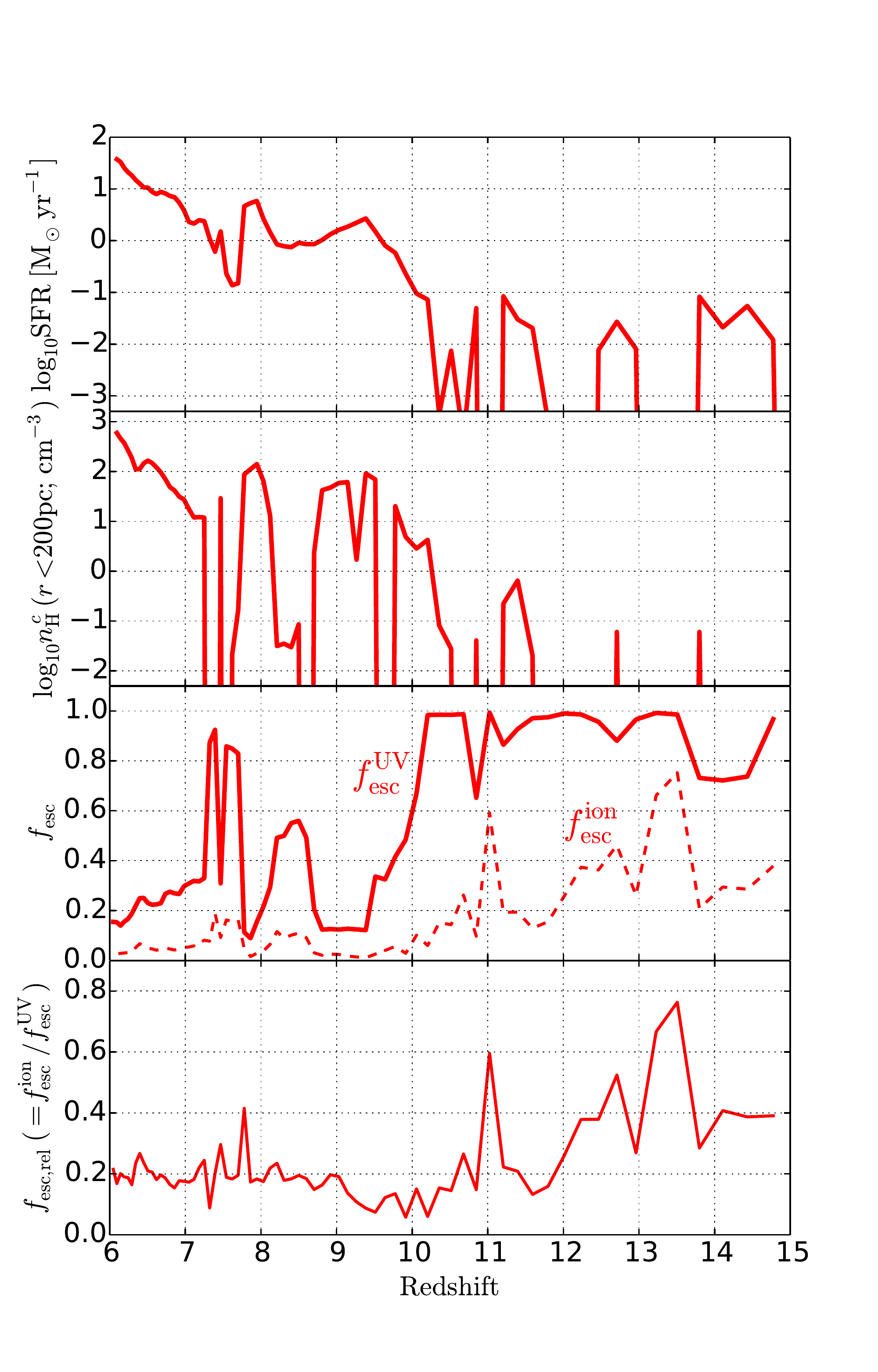}
\caption{
Redshift evolution of SFR (top panel), mean gas density within $200$~{\rm pc} from the galactic center (second panel). 
The third panel shows the absolute escape fraction for our fiducial run (Halo-11), where solid and dashed lines represent UV continuum ($1500-2800$~\AA) and Lyman-continuum photons ($\le 912$~\AA), respectively.
The bottom panel shows the relative escape fraction as defined in the main text. 
}
\label{fig:Halo-11}
\end{center}
\end{figure}

The upper and middle panels of Figure~\ref{fig:Halo-11} show the redshift evolution of SFR and  gas density at the galactic center of Halo-11 at $z = 6-15$.
Since Y17 already discussed them, here we explain only the important points briefly.
The figure shows the intermittent star formation of Halo-11, which is driven by the cycle of following processes:
(1) the central density increases as the gas is accreted due to the gravitational force of the halo, which drives active star formation;
(2) at the end of the lifetime of massive stars ($\sim 10~{\rm Myr}$), Type-II SNe occur, and the galactic wind is launched, resulting in the quenching of star formation.

The bottom panel of Fig.~\ref{fig:Halo-11} shows the evolution of  absolute escape fraction, which we define as follows:
\begin{equation}
\label{eq:escape}
f_{\rm esc,abs} \equiv \frac{ L_{\rm out}^{\rm UV}}{ L_{\rm int}^{\rm UV}},
\end{equation}
where $L_{\rm int}^{\rm UV},~L_{\rm out}^{\rm UV}$ are the intrinsic  and  emergent UV luminosities measured at the virial radius, respectively. 
We estimate the escape fractions of ionizing photons ($f_{\rm esc,abs}^{\rm ion}$) and non-ionizing UV continuum  ($f_{\rm esc,abs}^{\rm UV}$) for the wavelength range of $1500-2800$\,\AA.
We find that $f_{\rm esc,abs}^{\rm UV}$ fluctuates in the range of  $\sim 0.2-0.8$ at $z\lesssim 10$, but with $f_{\rm esc,abs}^{\rm UV}\gtrsim 0.8$ at $z>10$. 
As the gas density ($n_{\rm H}^{\rm c}$) at the galactic center increases, dusty clouds efficiently absorb UV photons, resulting in lower escape fraction.  
The time-averaged value is $\langle f_{\rm esc,abs}^{\rm UV} \rangle \sim 0.3$ at $z= 6-10$. 
Also, $f_{\rm esc,abs}^{\rm ion}$ rapidly changes in the range of $\sim 0.01-0.2$, with a time-averaged value of $\left<f_{\rm esc,abs}^{\rm ion}\right>\sim 0.06$ at $z= 6-10$. 

Observationally it is difficult to measure the absolute escape fraction directly, so instead the observers have defined the `relative' escape fraction which is easier to measure \citep[e.g.][]{Steidel01,Siana07,Vasei16}: 
\begin{equation}
\label{eq:escape}
f_{\rm esc,rel} \equiv \frac{f_{\rm esc,abs}^{\rm ion} }{f_{\rm esc,abs}^{\rm UV} } = \frac{(F_{\rm UV}/F_{\rm ion})_{\rm int}}{(F_{\rm UV}/F_{\rm ion})_{\rm obs}}{\rm e^{\tau_{\rm IGM, ion}}},
\end{equation}
where $F_{\rm UV}/F_{\rm ion}$ is the ratio of flux densities at non-ionizing UV continuum and ionizing photons. 
Observational studies have measured this ratio and estimated the relative escape fraction using the estimates from population synthesis models and IGM modeling, e.g., $(f_{\rm esc,abs}^{\rm ion} / f_{\rm esc,abs}^{\rm UV}) \simeq 0.7$ and  $\exp (- \tau_{\rm IGM, ion}) \simeq 0.4$ \citep{Siana07,Vasei16}. 
At $z < 2$, only several Lyman-continuum emitters have been observed \citep{Leitet11,Leitet13,Borthakur14,Izotov16},  typically with very low values of $f_{\rm esc,rel} \lesssim 0.1$.  
 At $2<z<4$, only a few robust detections have been made by 
 \cite{Vanzella12,Vanzella15,Vanzella16,Mostardi15,deBarros15} using high-resolution HST images.  
In addition, \cite{Micheva17} observed 25 candidates of Lyman-continuum emitters at $z=3.1$ using a narrow-band filter designed to detect ionizing photons at $\sim 900$\,{\AA} in the rest frame \citep[see also][]{Iwata09} and showed that $f_{\rm esc, rel}$ increased as the UV flux decreased. 

Here we show $f_{\rm esc, rel}$ by measuring the ratio $f_{\rm esc,abs}^{\rm ion}/f_{\rm esc,abs}^{\rm UV}$ as in the bottom panel of Fig.~\ref{fig:Halo-11}.
We find that $f_{\rm esc, rel}$ fluctuates in the range of $\sim 0.2 - 0.8$ at $z \gtrsim 10$ and becomes roughly constant with $\sim 0.2$ at $z < 10$.
The transition of $f_{\rm esc, rel}$ from higher to lower values as galaxies become brighter is similar to the trend observed in \cite{Micheva17}. 
In our simulations, ionizing photons can easily escape from galaxies at $z \gtrsim 10$ because most gas can be ejected due to the feedback. 
However, as the halo mass increases, neutral hydrogen in high-density gas clouds covers star-forming regions and prevents the escape of ionizing photons \citep[see also,][]{Yajima11,Yajima14}. 
The absorption by neutral hydrogen reduces $f_{\rm esc,abs}^{\rm ion}$ more substantially than $f_{\rm esc,abs}^{\rm UV}$, because only the dust absorption is relevant for $f_{\rm esc,abs}^{\rm UV}$. 

In addition, note that $f_{\rm esc, rel}$ is greater than $f_{\rm esc, abs}^{\rm ion}$ at $z \lesssim 10$. This is because  $f_{\rm esc, rel}$ does not take into account the dust extinction. As shown in the third panel of Fig.~\ref{fig:Halo-11}, more than half of UV continuum photons are absorbed by dust at $z \sim 6-10$. 
Therefore we suggest that the dust correction (conversion from $f_{\rm esc, rel}$ to $f_{\rm esc,abs}^{\rm ion}$) is required even for massive galaxies at high redshifts.

\begin{figure}
\begin{center}
\includegraphics[width=\columnwidth]{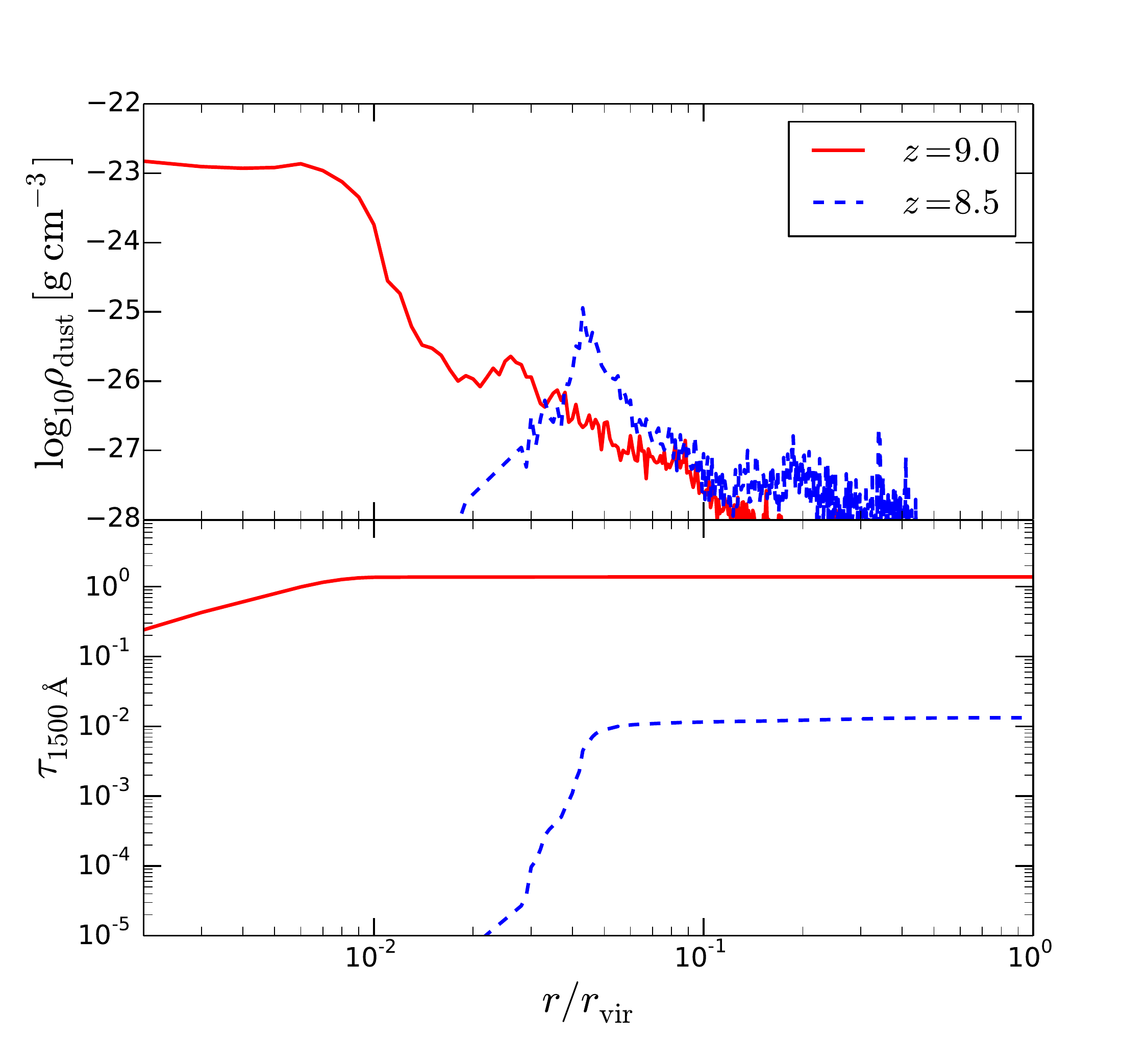}
\caption{
{\it Top panel}: 
Radial density profile of dust in Halo-11,  measured from the peak position of local SFR.
The dust density is computed by taking the mean value in each spherical shell with a width of $16$\,pc. 
{\it Bottom panel}: 
Optical depth for UV radiation ($1500$\,{\rm \AA}).
Red solid and blue dashed lines represent the snapshots when the UV escape fraction is low ($z=9.0$) and high ($z=8.5$), respectively. 
}
\label{fig:tau}
\end{center}
\end{figure}

The fluctuation of escape fraction in high-$z$ dwarf galaxies ($\Mh \sim 10^8-10^{10.5}~\Msun$) was discussed in previous works \citep{Kimm14,Paardekooper15,Trebitsch17}, although they focused on ionizing photons alone.
We find that escape fraction becomes very small when the galaxy becomes massive ($\Mh\sim 10^{10}-10^{11}~\Msun$), 
since most UV photons are absorbed by the central dense gas and dust.

To investigate the cause of fluctuating UV escape fraction, we plot 
the density distribution of dust and UV optical depths at $z=8.5$ \& $9.0$  in Figure~\ref{fig:tau}.
We see that $f_{\rm esc}^{\rm UV}$ is high (0.56) at $z=8.5$ and becomes low (0.12) at $z=9.0$. 
The horizontal axis is the distance from the peak position of local SFR, normalized by the virial radius.
The dust density profile is obtained by taking the mean values in each spherical shell.
As shown in the figure, the dust is concentrated in the star-forming region at $z=9.0$, which makes it optically thick.
At $z=8.5$, dust and gas are ejected due to SN feedback, and the star-forming region becomes optically thin.
Thus, we conclude that the time variation of $f_{\rm esc}^{\rm UV}$ is due to the cycle of gas accretion and galactic outflow.

In addition, we consider the time-scale of the transitions based on a simple spherical shell model.
At $z<10$, the total SNe energy released from a single starburst is
$\sim 10^{55} \left(\frac{SFR}{0.1~{\Msun~{\rm yr^{-1}}}}\right) \left(\frac{t_{\rm life}}{10~{\rm Myr}}\right)~{\rm erg} $,
where $t_{\rm life}$ is the lifetime of massive stars and we consider the energy of each supernova as $10^{51}~{\rm erg}$. 
It easily exceeds the gravitational binding energy of a star-forming gas cloud  $E_{\rm b,g}\sim GM_{\rm Jeans}^{2}/2L_{\rm Jeans}\sim 10^{53}~{\rm erg}$, as described in Y17 considering the Jeans instability. 
On the other hand, it is difficult to exceed the binding energy of the host halo, 
$E_{\rm b,h}
\sim 10^{56} \left( \frac{\Mh}{10^{10}~\Msun} \right)^{5/3} \left(\frac{1+z}{10}\right)
\left(\frac{\Omega_{\rm b}/\Omega_{\rm m}}{0.16}\right) ~{\rm erg}$. 
Thus, SNe can eject the gas from the central region but not from the halo.
Therefore, the outflowing gas shell will stall at a specific radius and fall back to the center within a free-fall time. 
The free-fall time for a shell to fall from the virial radius to the center is 
$t_{\rm ff} \sim \left( 2 r_{\rm vir}^{3} / G\Mh \right)^{1/2}\sim 100 \left(\frac{1+z}{10}\right)^{-3/2}~{\rm Myr}$.
This is consistent with simulation results because the fluctuations occur a few times within $450~{\rm Myr} ~(z=6-10)$.

\subsection{Sub-millimeter flux and halo mass dependence}
\label{sec:flux}

As $f_{\rm esc}^{\rm UV}$ rapidly changes, galaxies are expected to flicker in both UV and sub-mm wavelengths on short time-scales. 
Here we investigate the time evolution of sub-mm flux. 
Since dust extinction is likely to depend on the halo mass \citep[e.g.,][]{Yajima11, Yajima14}, 
we compare the sub-mm fluxes of different halos. 

\begin{figure}
\begin{center}
\includegraphics[width=\columnwidth]{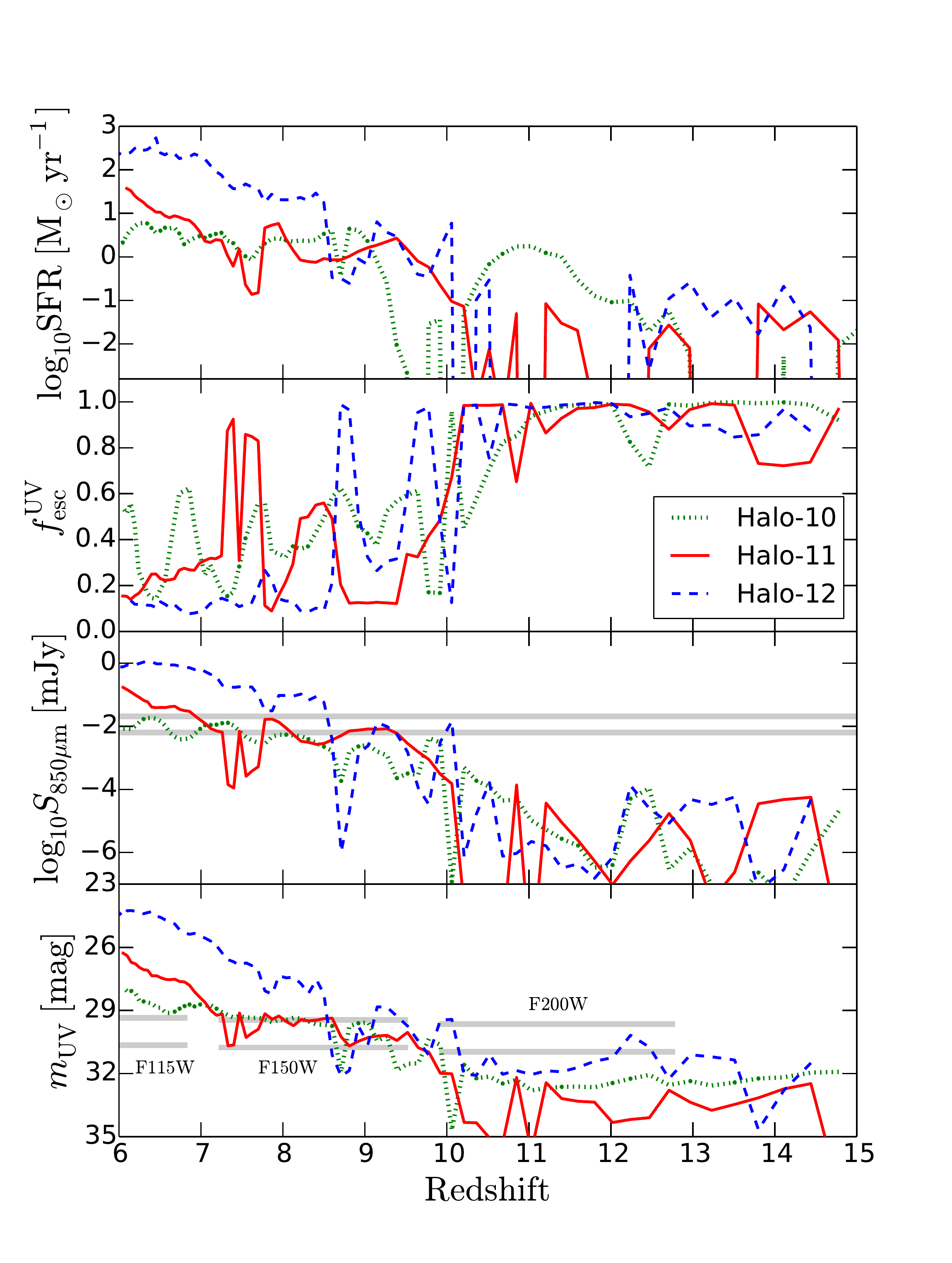}
\caption{
Redshift evolution of SFR (top panel), 
UV escape fraction (second panel), sub-mm flux (third panel) and emergent UV flux (bottom panel) are shown for Halo-10 (green dotted), Halo-11 (red solid), and Halo-12 (blue dashed). 
The gray horizontal lines in the third and forth panels show 
$3\sigma$ (lower line) and $10\sigma$ (higher line) detection thresholds for ALMA (full operation) and JWST with 10 hours time integration, which we obtained from the online calculators of each observatory. 
}
\label{fig:Mhalo}
\end{center}
\end{figure}

The top panel of Figure~\ref{fig:Mhalo} shows the SF histories of different haloes. 
All haloes show gradual increase in SFR with decreasing redshift, but with intermittent star formation at $z \gtrsim 10$ which was also discussed in Y17.
The second panel shows the rapid fluctuation of  $f_{\rm esc}^{\rm UV}$ at  $z\gtrsim 10$ together with star formation. 
As the dust mass increases with galaxy growth, the massive galaxies have lower  $f_{\rm esc}^{\rm UV}$ at $z < 10$:
 $f_{\rm esc}^{\rm UV} = 0.1 - 0.2$ for  Halo-11 (Halo-12) at $z \lesssim 7.5\;(8.5)$, while $f_{\rm esc}^{\rm UV}$ of Halo-10 fluctuates in the range of $\sim 0.2 - 0.6$ even at $z \sim 6$. 
In the case of Halo-12, the  baryon mass within 200\,pc from the galactic center becomes $\sim 10^{10}~\Msun$ at $z\sim 8.5$, and the gravitational binding energy exceeds the thermal energy released by SNe  in an intense star formation  with ${\rm SFR} \gtrsim 10~\Msun~{\rm yr^{-1}}$. 
Therefore the gas is able to stay around star-forming regions and obscure UV light. 
These variations of $f_{\rm esc}^{\rm UV}$ for different halo masses are closely linked to the UV luminosity functions.  
For example,  if the UV magnitude of $m_{\rm UV}\sim 30$ fluctuates by a factor of 2 at $z\sim 7-8$, it can make the luminosity function flatter at the faint end. 
In our future work, we will investigate the impact  of fluctuating escape fraction on the luminosity function using large-scale cosmological simulations. 

The third panel of Fig.~\ref{fig:Mhalo} shows the redshift evolution of sub-mm flux  at $850~{\rm \mu m}$ in the observed frame.
In all cases, the flux fluctuates by more than an order of magnitude at $z\lesssim 10$ because of the starbursts and low UV escape fractions. 
When the dusty gas is concentrated at the galactic center, it causes a starburst and obscures the UV light from young stars, resulting in high sub-mm fluxes. 
The sub-mm flux of Halo-11 (Halo-12) reaches $S_{\rm 850\mu m}^{\rm obs.} \sim 0.1\,{\rm mJy} ~(1\,{\rm mJy})$ at $z\sim 6-7$, which can be observed by ALMA with $\sim$ 3-hour integration time. 
As a reference, we compare our models with the sensitivity of ALMA 10-hour observation as shown in the figure.
The Halo-11 is detectable at $z\lesssim 7.3$ or $z\sim 8.0,~9.0$ with $3\sigma$ significance.
We can compare Halo-11 and Halo-12 with one of the high-$z$ galaxies observed by ALMA, A1689-zD1 at $z\approx 7.5$ \citep{Watson15}.
The sub-mm flux of the galaxy was $0.61\pm 0.12~{\rm mJy}$, and the estimated SFR and dust mass were $\sim 12~\Msun~{\rm yr^{-1}}$ and $\sim 4\times 10^{7}~\Msun$.
The A1689-zD1 is a sub-$L_{\rm *}$ galaxy, thus it was considered as one of the `normal' galaxies which are the dominant population at that epoch.
The dust masses of Halo-11 and Halo-12 at $z=7.5$ are $\sim 5\times 10^{6}~\Msun$ and $\sim 8\times 10^{7}~\Msun$, respectively;  thus A1689-zD1 has intermediate parameters of Halo-11 and Halo-12. 

In addition to A1689-zD1, the LBGs with sub-mm fluxes of $\sim 0.1\,{\rm mJy}$ have been  detected by recent ALMA observations \citep{Bowler18,Hashimoto18b,Tamura18}.
Halo-11 successfully reproduces the observed  sub-mm and UV fluxes of LBGs. 
However, note that the observations which detect dust continuum emission at only one wavelength has to assume the dust temperature to estimate the SFR or dust mass.
If this assumption is not valid, then the estimated physical values could be incorrect.
In Sec.~\ref{sec:dust}, we discuss the typical dust temperature in the first galaxies.

The bottom panel of Fig.~\ref{fig:Mhalo} shows the evolution of apparent UV ($1500$\,{\AA}) magnitude.
As a reference, we also show the sensitivity of JWST in the figure.  
This comparison suggests that the detection of Halo-10 and Halo-11 at $z>10$ is very difficult, 
whereas the $\lya$ flux from the first galaxies can be detected even at $z \gtrsim 10$ \citep{Yajima15}.
\cite{Yajima18} showed that the massive galaxies tend to have higher SFR, and the strong ionizing flux forms huge H{\sc ii} bubbles ($\gtrsim 100~{\rm kpc}$) which increases the IGM transmission of $\lya$ photons.

\subsection{Duty cycle and Observability}
\label{sec:duty}

As we discussed in Sec. \ref{sec:flux}, the sub-mm brightness of the first galaxies oscillates on a short time-scale. 
Hereafter we call this phenomena `{\it duty cycle}' ($f_{\rm duty}$), and attempt to estimate its value. 
The duration of the bright phases is directly related to the detection probability of galaxies with a specific mass.
\citet{Jaacks12b} studied SF histories of galaxies at $z>6$ using cosmological hydrodynamic simulations, and estimated the duty cycle defined as the ratio of the number of galaxies which is brighter than the detection limit of the HST to the total number.

In this work, we resolve the spatial distribution of dust in detail by using the zoom-in initial conditions and calculating the dust attenuation.
Here we estimate $f_{\rm duty}$ based on the detectability of sub-mm fluxes of galaxies by ALMA.
We define $f_{\rm duty}$ as the ratio of number of galaxies brighter than the detection limit ($S_{\rm th}$) at $850~{\rm \mu m}$ to the total number:
\begin{equation}
f_{\rm duty} \equiv \frac{N(S> S_{\rm th})}{N_{\rm tot}}.
\end{equation}

To increase the number of galaxies, we calculate the radiative properties of all satellite galaxies in the zoom-in regions, and include them in our sample. 
The mass limits of the satellite galaxy selection are $\Mh > 10^{8}~\Msun$ for Halo-11 and $\Mh > 10^{9}~\Msun$ for Halo-12, respectively.

\begin{figure}
\begin{center}
\includegraphics[width=\columnwidth]{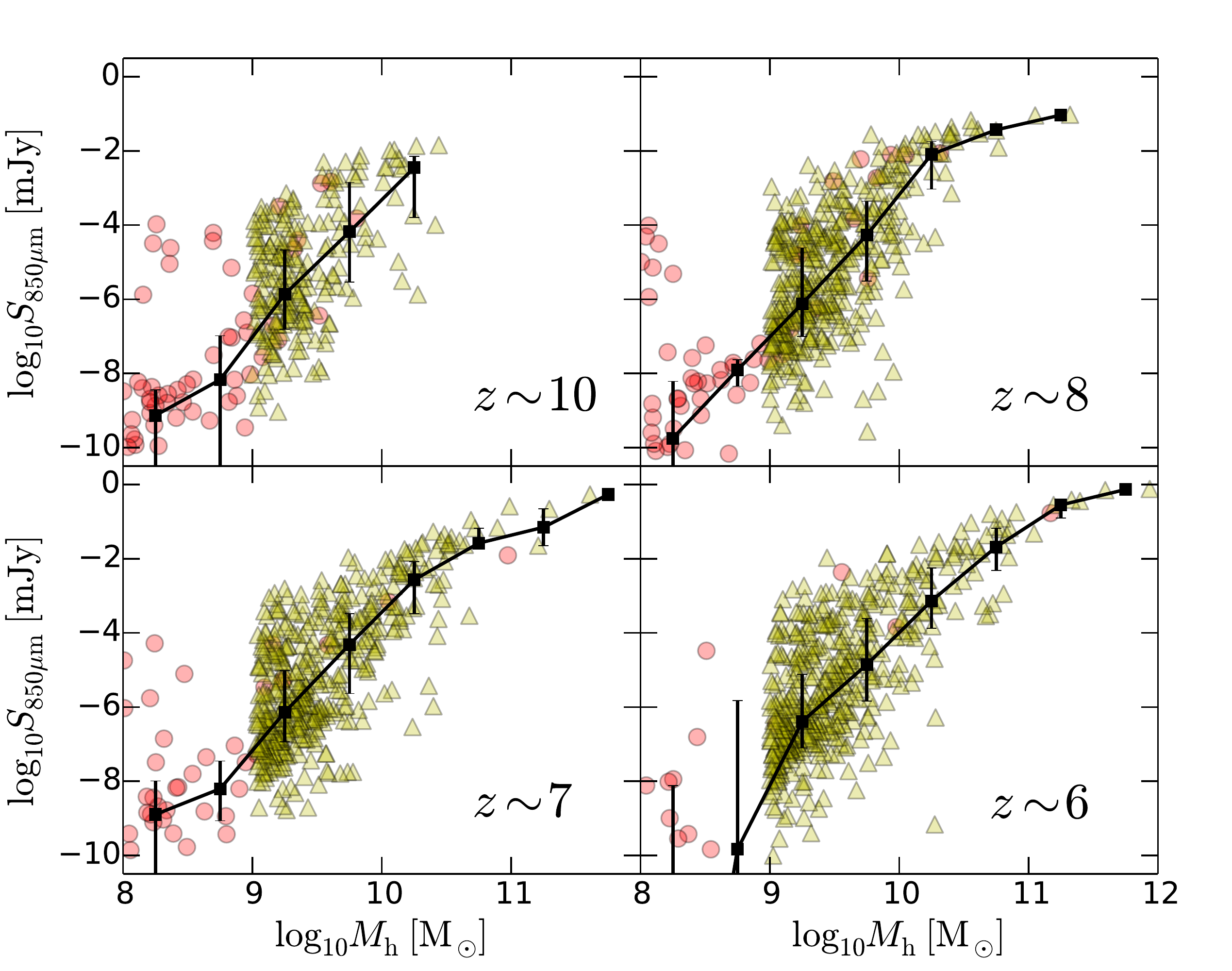}
\caption{
Sub-mm fluxes of all main and satellite galaxies in Halo-11 (red circles) and Halo-12 (yellow triangles) at $z\sim 10,~8,~7,$ \& $6$.
Black squares are the median values in each halo mass bin with the width of $\Delta \log_{10}{(\Mh/\Msun)}=0.5$, and the error bars show the quartiles.
}
\label{fig:flux_dist}
\end{center}
\end{figure}

Figure~\ref{fig:flux_dist} shows the distribution of the halo masses and the sub-mm fluxes of our samples at $z\sim 10,~8,~7$, \& $6$.
With decreasing redshift, one can see that the massive end of the distribution gradually shifts to the right-hand-side as the haloes grow in their masses.  The sharp cutoff in the distribution of yellow triangles are due to the halo mass resolution of Halo-12, which corresponds to roughly $10^{3}$ dark matter particles.  Halo-11 has higher resolution than Halo-11,  and haloes with  $M_h\gtrsim 10^8\,h^{-1}{\rm M}_\odot$ are resolved reasonably well. 
The sub-mm flux increases steeply with the halo mass because massive galaxies have higher SFRs and lower $f_{\rm esc}$ (see Sec.~\ref{sec:flux}).
Star formation of satellite galaxies also occurs intermittently, resulting in the large dispersion of the sub-mm fluxes. 

\begin{figure}
\begin{center}
\includegraphics[width=\columnwidth]{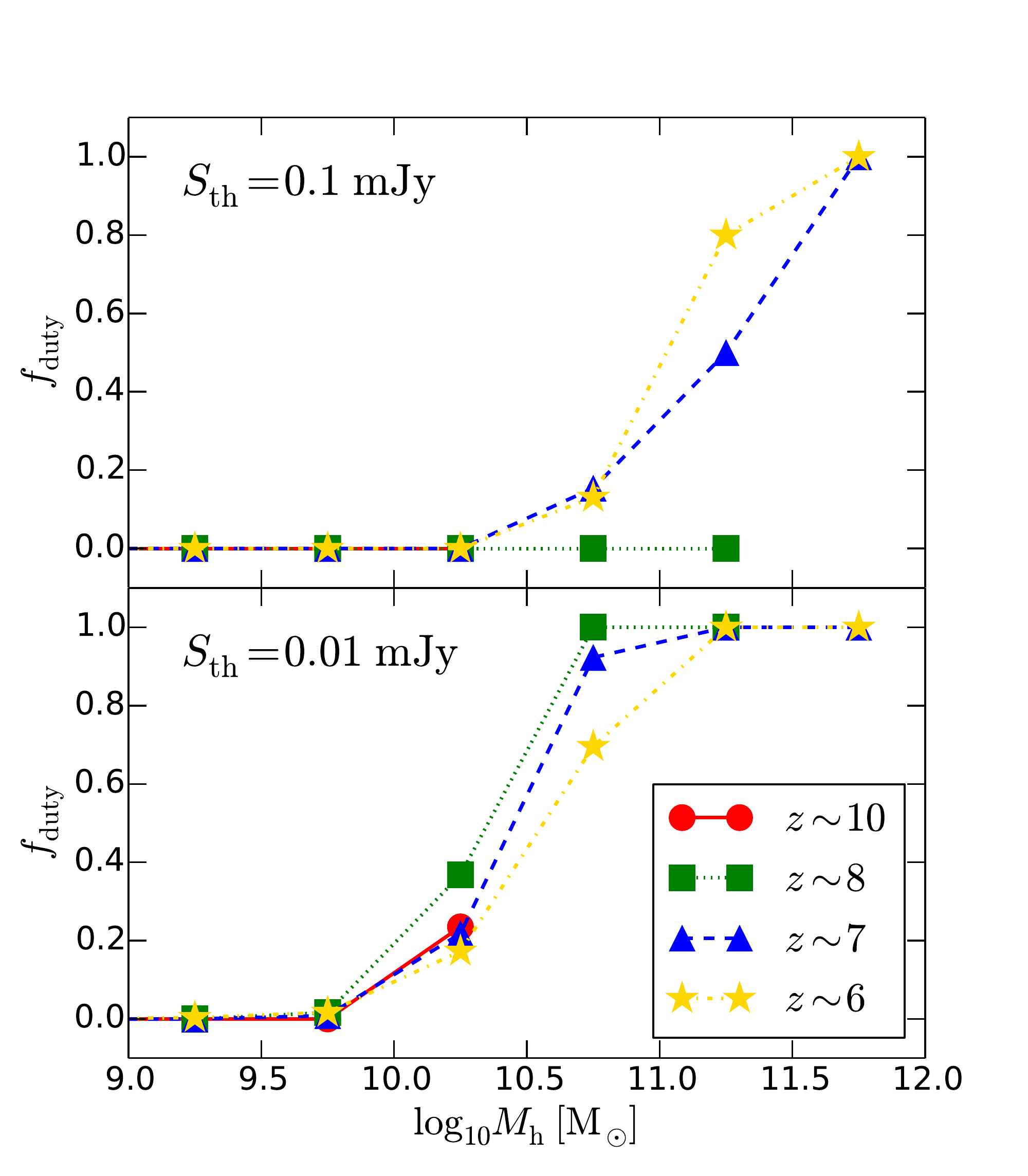}
\caption{Duty cycle ($\un{f}{duty}$), which is defined as the number fraction of observable galaxies to the total sample in each halo mass bin, is plotted against halo mass. 
The top panel shows $\un{f}{duty}$ for the detection threshold of $S_{\rm th}=0.1~{\rm mJy}$ at $850~\rm \mu m$.  This corresponds to $10\sigma$ detection by ALMA observation with 20 min time integration with full operation.   The bottom panel represents the case of $S_{\rm th}=0.01~{\rm mJy}$ which corresponds to $10\sigma$ detection with $40$ hours integration.
}
\label{fig:fduty}
\end{center}
\end{figure}

Figure~\ref{fig:fduty} shows $f_{\rm duty}$ as a function of halo mass. 
We separate the galaxies into halo mass bins  with the bin size of $\Delta\log_{\rm 10}(\Mh/\Msun) = 0.5$ and derive  $f_{\rm duty}$ for each bin.
Upper and lower panels present  $f_{\rm duty}$ for the detection thresholds of $S_{\rm th}=0.1$ and $ 0.01\,{\rm mJy}$, which corresponds to $10\sigma$ detections with 40 minutes and 60 hours observations by ALMA Band-7 full operation, respectively. 
In the case of $S_{\rm th}=0.1~{\rm mJy}$,  $f_{\rm duty}$ exceeds 0.5 only for massive halos with  $\log_{\rm 10}(\Mh/\Msun) \geq 11$ at $z\leq 7$, and it  changes with redshift only at the massive end. 
Note that, however, the sample at $M_{\rm h} > 10^{11}~\Msun$ is small ($N \lesssim 10$). 
Therefore the estimation of $f_{\rm duty}$ at the massive end is likely to contain large uncertainties. 
In the case of $S_{\rm th}=0.01~{\rm mJy}$,  $f_{\rm duty}$ steeply increases at lower halo mass and becomes 0.5 at $\log_{\rm 10}(\Mh/\Msun) \approx 10.5$. 
Therefore deep observations that can observe down to $0.01~{\rm mJy}$ will be able to detect sub-mm flux from low-mass galaxies.

\begin{figure}
\begin{center}
\includegraphics[width=\columnwidth]{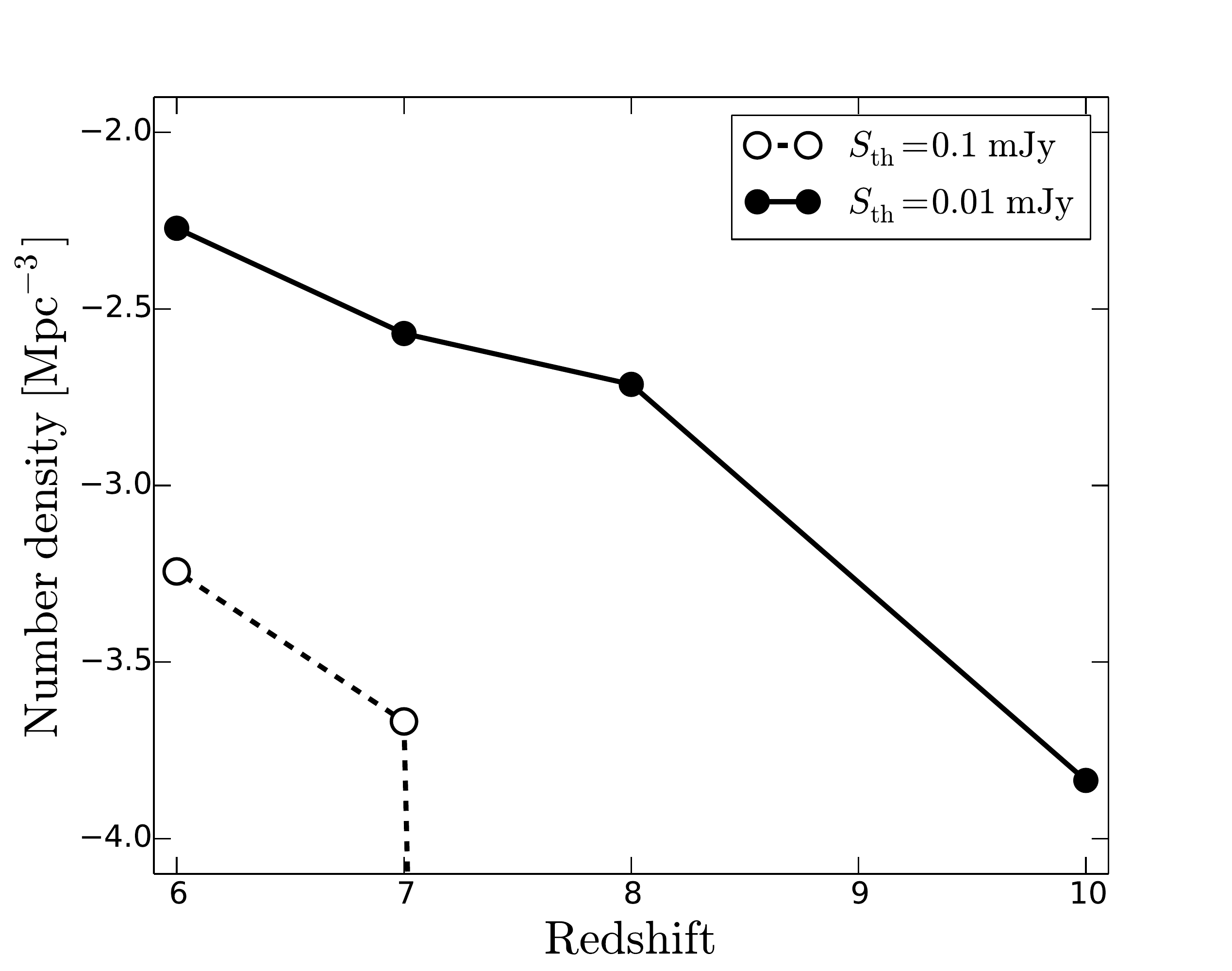}
\caption{Number density of observable sub-mm sources in units of comoving ${\rm Mpc}^{-3}$, which is  estimated by integrating the Sheth \& Tormen (2002) halo mass function multiplied by the duty cycle.  The solid and dashed lines represent the predicted number densities for the detection thresholds of $S_{\rm th}=0.01$ and $0.1~{\rm mJy}$ at $850~{\rm \mu m}$, respectively.}
\label{fig:numdens}
\end{center}
\end{figure}

By integrating the \citet{Sheth02} halo mass function multiplied by  $f_{\rm duty}$, we roughly estimate the number density of observable sub-mm sources as shown in Figure~\ref{fig:numdens}. 
In the case of $S_{\rm th}=0.01~{\rm mJy}$, the number density increases as redshift decreases, and it becomes $5.4\times 10^{-3}~{\rm cMpc^{-3}}$ at $z\sim 6$.
This number density is close to that of LBGs at $z \sim 6$ \citep[$\sim 10^{-2} ~{\rm cMpc^{-3}}$;][]{Ouchi10, Bouwens15}.
Therefore, we suggest that the deep sub-mm observations allow us to detect even typical LBGs. 
On the other hand, observations with $S_{\rm th}=0.1\,{\rm mJy}$ won't be able to detect typical LBGs.

\begin{figure}
\begin{center}
\includegraphics[width=\columnwidth]{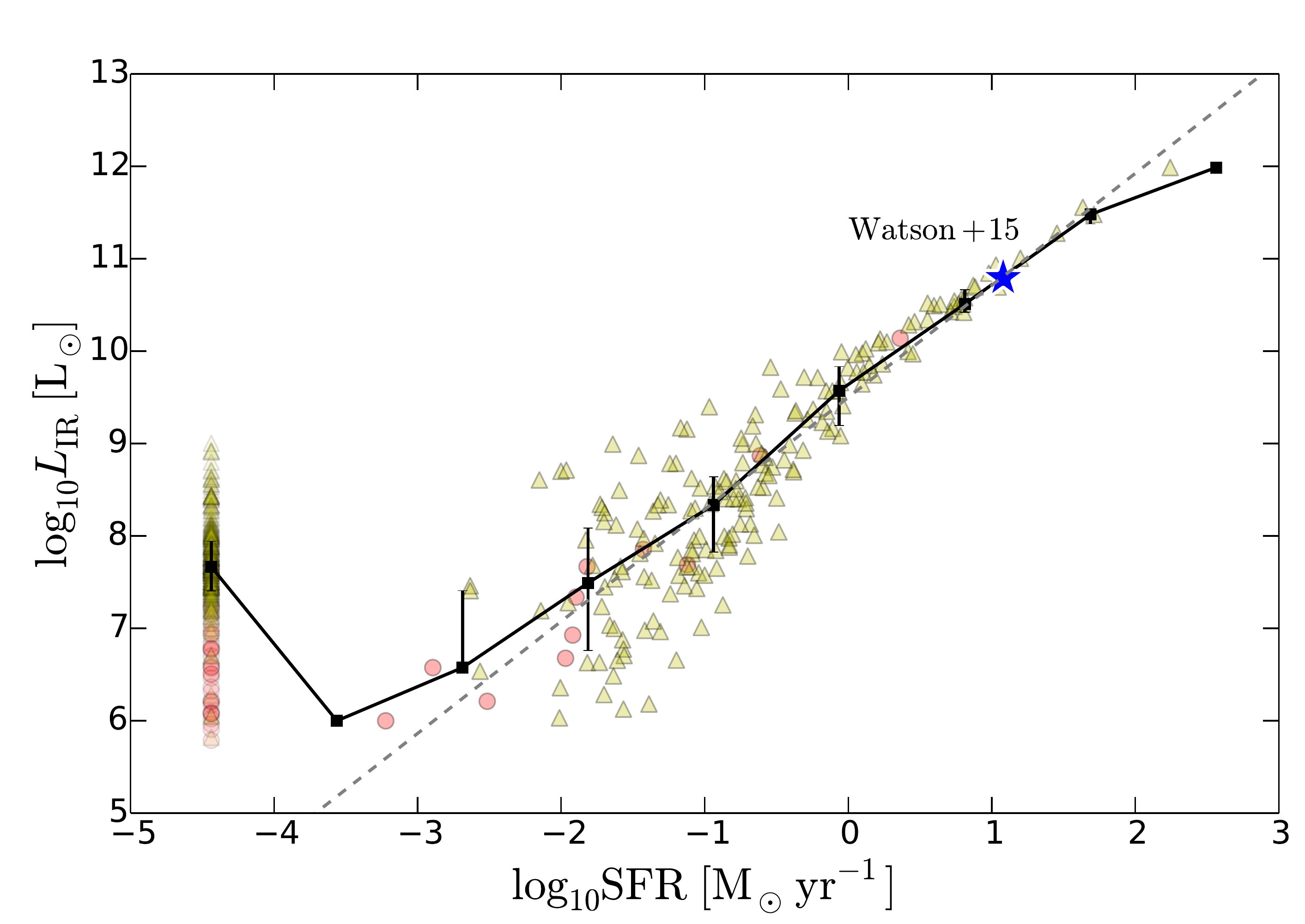}
\caption{Relation between SFR and IR luminosity at $z= 7$.  Meaning of the symbols is the same as in Fig.~\ref{fig:flux_dist}. Galaxies with zero SFR are shown at $\log_{10}{{\rm SFR}}=-4.5$ for plotting purposes.  The dashed line shows the power-law fit (Eq.~\ref{eq:fit}) for star-forming galaxies. Blue star represents an observed LBG at $z\approx 7.5$ \citep[A1689-zD1,][]{Watson15}.
}
\label{fig:sfr_vs_LIR}
\end{center}
\end{figure}

In addition, 
Figure~\ref{fig:sfr_vs_LIR} presents the relation between SFR and IR luminosity at $z=7$. For star-forming galaxies, the IR luminosity increase with SFR because young stars  dominantly generate the UV photons which are absorbed and reprocessed by dust.
The power-law fit of the relation follows as
\begin{equation}
L_{\rm IR} \approx 3.2\times 10^{9}\,{\rm \Lsun} \left( \frac{\rm SFR}{\rm \Msun\,yr^{-1}}\right)^{1.21}
\label{eq:fit}
\end{equation}
where we used the galaxy sample with  $-4<\log_{10}{\rm SFR~[\Msun~yr^{-1}]}<3$ for the fitting.  
If the dust completely absorbs UV photons, the power index must be unity because UV luminosity relates to SFR linearly. In our simulation, however, the low-mass galaxies have low SFR and high escape fraction of UV photons, resulting in low IR luminosity. We find that this effect makes the relation steeper.  
The relation can reproduce the observed LBG at $z\approx 7.5$ \citep{Watson15} remarkably well, whose SFR was measured from the UV and IR luminosity.
Meanwhile, non-star-forming galaxies also have large dispersion of IR luminosity due to the dust absorption of UV photons emitted by residual young stars.

\begin{figure}
\begin{center}
\includegraphics[width=\columnwidth]{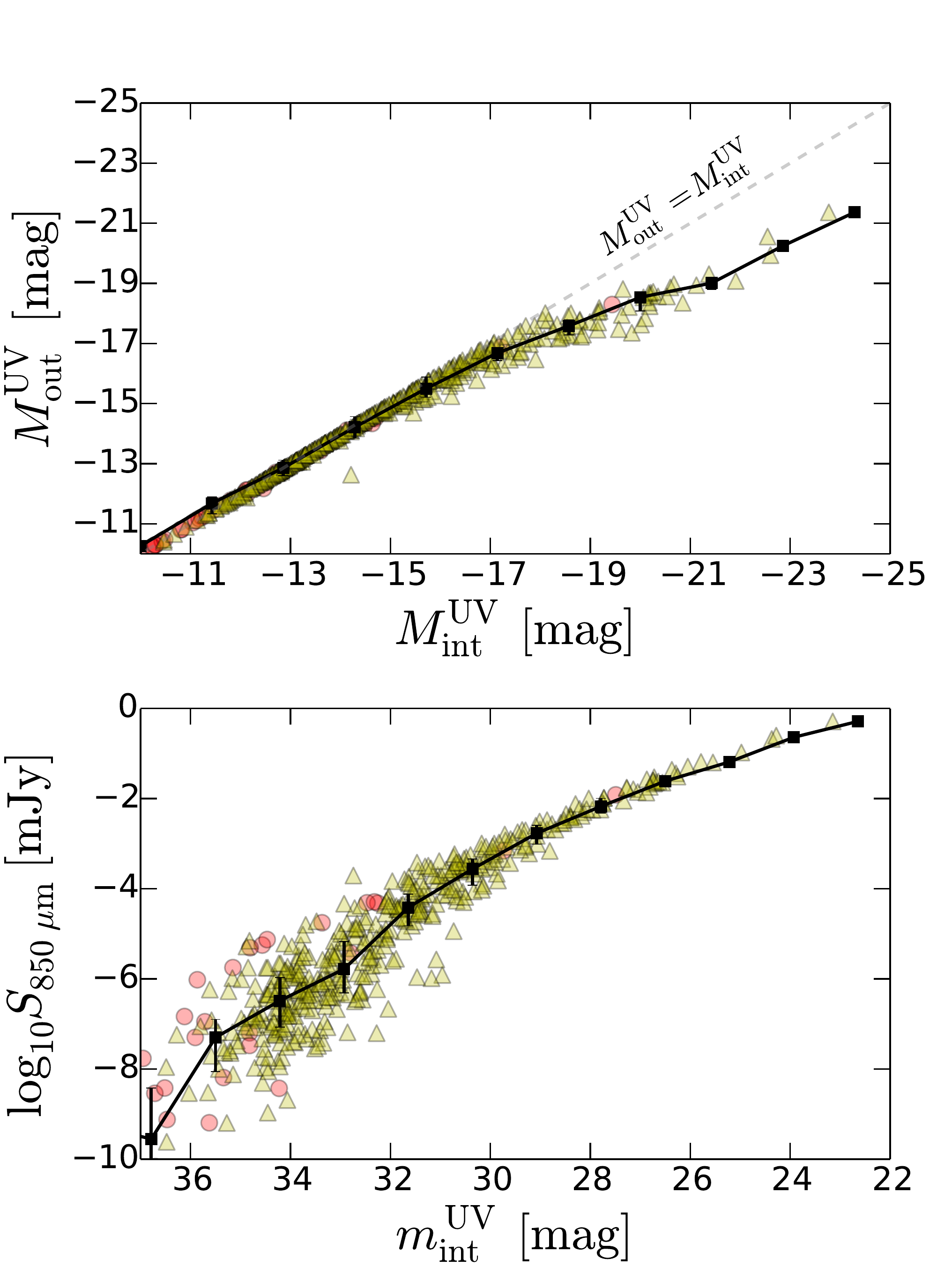}
\caption{Upper panel: Relation between intrinsic rest-frame UV absolute magnitudes and emergent ones of the main halo and satellites in Halo-11 (red circle) and Halo-12 (yellow triangle) at $z\sim 7$.  The black squares and error-bars show the medians and quartiles within each bin, respectively.  The dashed line describes the case in which all of UV photons can escape from the halo.  One can see that the dust attenuation strongly affects the UV magnitude by $\Delta M \sim 2~{\rm mag}$ for the bright galaxies. 
Lower panel:  Relation between the intrinsic UV apparent magnitude and observed sub-mm flux of simulated galaxies .}
\label{fig:ratio}
\end{center}
\end{figure}

Here we overview how  dust attenuation affects observation.
The top panel of Figure~\ref{fig:ratio} presents the relation between the intrinsic UV absolute magnitude and the emergent one, i.e., before and after the dust absorption in each galaxy.  
For the faint galaxies dimmer than $M_{\rm int}^{\rm UV}\sim -17$, most of the UV photons escape from the haloes, resulting in a tight linear relation between the emergent luminosities and the  intrinsic ones.
For the galaxies brighter than  $M_{\rm int}^{\rm UV}\sim -19$, the dust attenuation significantly decreases the UV luminosity (escape fraction $\sim 0.1$).

The bottom panel of Figure~\ref{fig:ratio}  shows the relation between the intrinsic apparent magnitude and the sub-mm flux in the observed frame.
The observed sub-mm flux decreases as $m_{\rm int}^{\rm UV}$ becomes fainter, and they reach the detection limit of ALMA ($\gtrsim 0.01~{\rm mJy}$) at $m_{\rm int}^{\rm UV}\sim 27$.
For the galaxies fainter than $m_{\rm int}^{\rm UV} \sim 29$\,mag,  the escape fraction rapidly changes between $0.2-0.8$ due to intermittent star formation, resulting in  dispersion of $\Delta m \sim 1.5$ for the observed magnitude.  

%
%

\subsection{Dust temperature}
\label{sec:dust}

The dust temperature is an important factor in estimating the bolometric infrared luminosity, but it has not been estimated well for the first galaxies so far due to limited observational data. 
In infrared observations of  local galaxies, the flux at some different frequencies can be detected, and the dust temperature is determined from the peak wavelength of modified black-body spectrum \citep[e.g][]{Hwang10}.
On the other hand, most of the ALMA observations of high-$z$ galaxies have obtained the flux only at one wavelength.
Therefore  the dust mass or SFR have been estimated based on the assumed dust temperature \citep[e.g. $\sim 40~{\rm K}$,][]{Watson15}.
Here we investigate the typical dust temperature of first galaxies.

To describe the spatial distribution of dust temperature, we present the 2-D map in the lower left panel of Fig.~\ref{fig:maps}, which displays the mass-weighted mean temperature along the line of sight.
In the outer parts of the galaxy,  cool dust ($T_{\rm d}\sim 30~{\rm K}$) dominates, while at the central star-forming region dust temperature is high ($T_{\rm d}\sim 60~{\rm K}$) due to strong UV irradiation. 

\begin{figure}
\begin{center}
\includegraphics[width=\columnwidth]{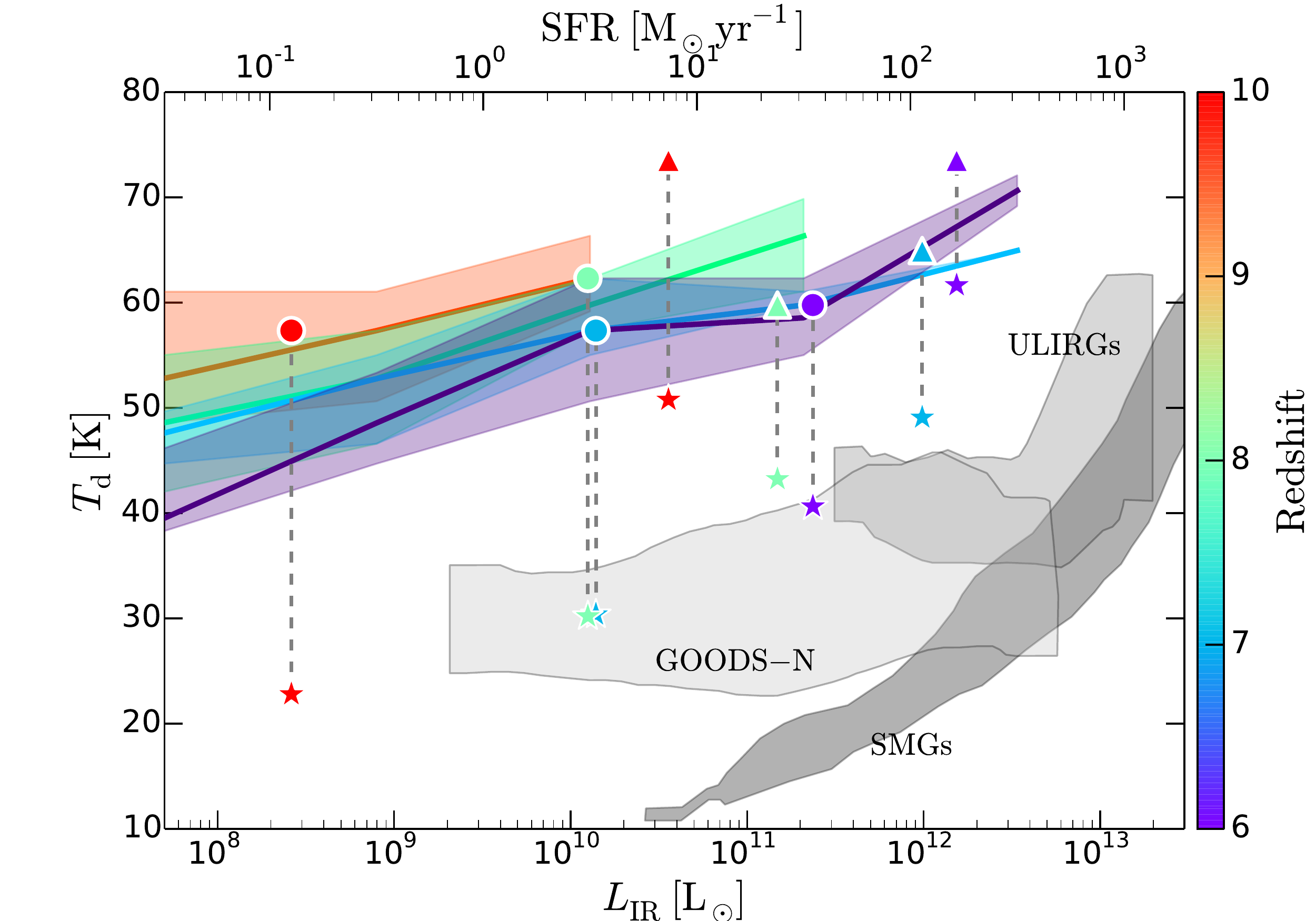}
\caption{
Relation between dust temperature and bolometric infrared luminosity. 
The upper horizontal axis indicates SFR derived from Eq.\,(\ref{eq:fit}).
Filled circles show the dust temperature measured from the peak wavelength of modified SEDs.
Different colors mean different redshifts: $z=6$ (purple), $z=7$ (cyan), $z=8$ (green), and $z=10$ (red).
Color-shaded regions indicate the ranges of dust temperatures of all satellite galaxies. 
Filled stars show the mean dust temperature weighted by dust mass of main haloes.
Gray shaded regions represent the range of dust temperatures of observed sub-mm galaxies (SMGs) at $z\sim 1-3$ \citep{Chapman05,Kovacs06}, ultra-luminous infrared galaxies (ULIRGs) at $z < 1$ \citep{Yang07,Younger09}, and typical star-forming galaxies at $z\sim 0.1-2.8$ \citep{Hwang10}. 
}
\label{fig:Td_sfr}
\end{center}
\end{figure}

Figure~\ref{fig:Td_sfr} shows the relation between dust temperature and total IR ($3-1000~{\rm \mu m}$) luminosity for the main and satellite galaxies in Halo-11 and Halo-12 at $z = 6-10$. 
We estimate the dust temperature from the peak wavelength of SEDs with modified black body spectrum considering the absorption efficiency of our dust model. 
The temperature increases from $\sim 40~{\rm K}$ to $\sim 70~{\rm K}$ as the IR luminosity increases from $\sim 10^{8}~\Lsun$ to $\sim 10^{12}~\Lsun$. 
Since high IR luminosity corresponds to intense star formation (Fig.~\ref{fig:sfr_vs_LIR}), the emitted UV photons heat dust efficiently.
For bright galaxies ($L_{\rm IR} > 10^{10}~\Lsun$) the temperatures are higher than that of nearby star-forming galaxies, local ULIRGs and low-$z$ SMGs (shaded areas, \citealt{Hwang10}) by about $10-20~{\rm K}$.
The compactness of high-$z$ galaxies induces formation of dense star-forming gas clumps and radiative heating of dust \citep[see also][]{Behrens18}.
In addition, Fig.~\ref{fig:Td_sfr} also displays mass-weighted mean dust temperatures by the star symbols, which are  lower than $T_{\rm peak}$ by $\sim 20-40~{\rm K}$.  This indicates that some fraction of dust is in cold state with $T \lesssim 30~\rm K$. 

\begin{figure}
\begin{center}
\includegraphics[width=\columnwidth]{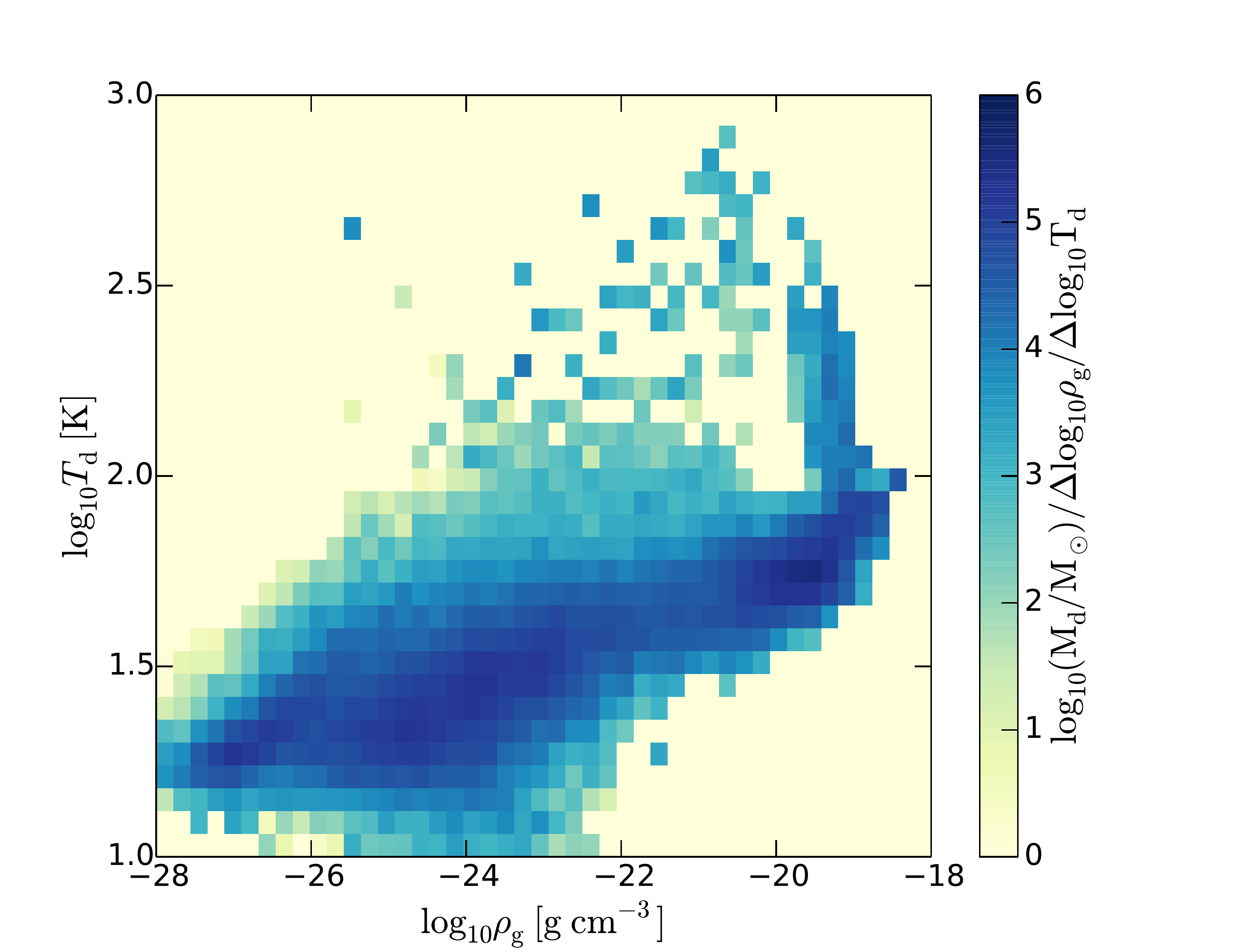}
\caption{
Distribution of dust mass as functions of gas density and dust temperature in Halo-11 at $z\sim 6$.
The cool dust ($\lesssim 40~{\rm K}$, $\sim 1.2\times 10^{7}~\Msun$) exists in the outer low-density regions, and the hot dust ($\gtrsim 40~{\rm K}$, $\sim 5\times 10^{6}~\Msun$) heated by stellar UV radiation exists in the central high-density regions.
}
\label{fig:phase}
\end{center}
\end{figure}

Figure~\ref{fig:phase} shows the distribution of dust mass as functions of  gas density and dust temperature.
We can see the evolution of dust temperature on the figure.  In the outer low-density regions,  a large amount of dust have a temperature similar to the CMB temperature ($\sim 30~{\rm K})$.  As dust is accreted onto the galactic center, dust temperature increases to $\sim 100~{\rm K}$ due to strong irradiation of stellar UV photons.
Observational studies of high-$z$ galaxies using flux of $850~{\rm \mu m}$ estimated the amount of central hot dust (see Fig.~\ref{fig:maps}), however, our simulation suggests that there is comparable amount of dust in the outer part of the halo.   The longer wavelength observation would be studying the distribution of cooler dust in the outer parts of high-$z$ galaxies. 

As redshift increases, galaxies become compact ($\sim 10$ per cent of virial radius, see Sec.~\ref{sec:size}). 
Therefore we expect that dust also distribute compactly, resulting in efficient heating of dust by intense stellar UV flux.
Here we estimate the typical distance of dusty clouds from star-forming regions under the assumption of radiative equilibrium as
\begin{equation}
R = \left(
\frac{ L_{\rm UV} }{ 16 \pi^{2} \int  Q_{\rm \nu}  B_{\rm \nu}(T_{\rm d}) d\nu }
\right)^{1/2},
\end{equation}
where $Q_{\rm \nu}$ is the absorption efficiency to the geometrical cross section of dust. 
Here we use $Q_{\rm \nu}$ estimated in \cite{Laor93}.  
In the case of our simulated dust temperature $T_{\rm d} \sim 50-70~{\rm K}$ and $L_{\rm UV}=L_{\rm IR}=10^{11}~\Lsun$, the typical distance is $1-3$\,kpc, 
which is similar to the disk sizes of high-$z$ galaxies. 
Note that, however, the dust temperature depends on not only the compactness, but also the size of dust grains.
\citet{Nozawa07} suggested that the dust size ranged $\gtrsim 0.1~\rm \mu m$ because 
small dust grains with $< 0.1~\rm \mu m$ could be destroyed due to reverse shocks in supernova remnants.
If we consider different dust model with larger grain sizes, the dust temperature will decrease \citep[e.g.,][]{Yajima14a, Yajima17b}.

%
%
\subsection{Size evolution}
\label{sec:size}

In this sub-section we investigate the time evolution of galaxy sizes.
In the classical picture, the disk size  is proportional to the angular momentum of accreted gas \citep{Mo98}.
However, \citet{Genel15} showed that the galactic angular momentum was redistributed due to stellar feedback in the Illustris simulation (see also \citealt{Scannapieco08,Zavala08}).
They found that the feedback processes change the galactic morphologies and reproduce the observational relation \citep{Fall13} between the specific angular momentum and stellar mass at $z=0$.
Here we discuss how the sizes of clumpy high-$z$ galaxies change with time and affect the detectability of galaxies.
Our simulated galaxies have disks at $z\lesssim 10$, and the sizes are affected by SN feedback as described in Y17.

\begin{figure}
\begin{center}
\includegraphics[width=\columnwidth]{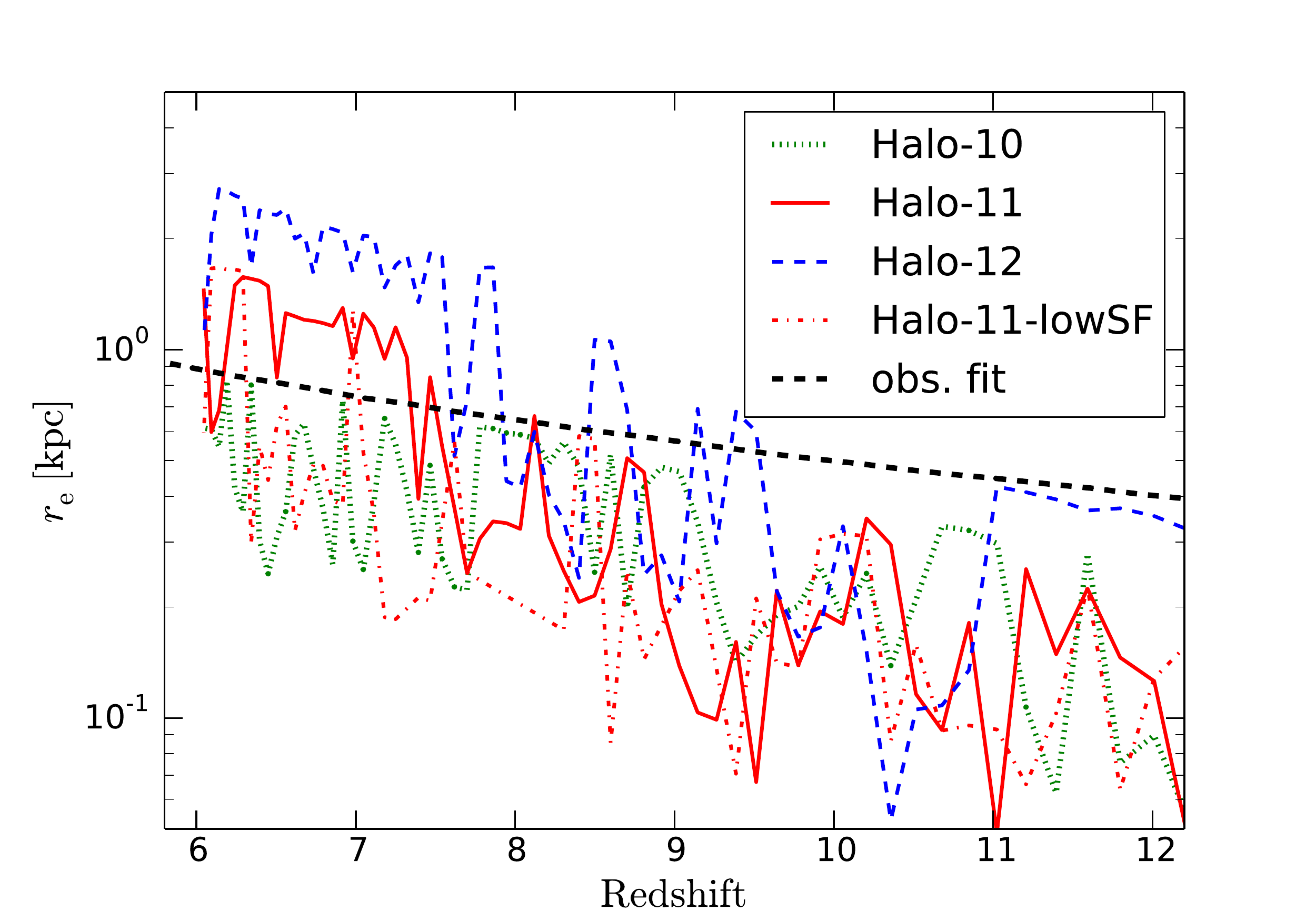}
\caption{
Redshift evolution of the half-light radius of the main galaxy in Halo-10 (green dotted), 11 (red solid), \& 12 (blue dashed), derived from the UV surface brightness distribution. 
Red dot-dashed line shows the half-light radius of Halo-11-lowSF. 
Black dashed line shows observational fit which use bright galaxies of $-21\lesssim M_{\rm UV}\lesssim -20.15$ (Kawamata et al. 2018).
}
\label{fig:re}
\end{center}
\end{figure}

Figure~\ref{fig:re} shows the UV half-light radii ($r_{\rm e}$) of Halo-10, Halo-11 and  Halo-12 as a function of redshift.
Here we define $r_{\rm e}$ (in physical kpc) as the distance from the brightest pixel within which the integrated brightness is half of total luminosity.
In the case of merging galaxies, companion galaxies may make $r_{\rm e}$ too large artificially. 
Therefore we set the upper limit of $r_{\rm e}$ as 10 per cent of virial radius.
We observe that $r_{\rm e}$ fluctuates around $0.02 - 0.1$ times the virial radius.
When the SFR becomes high, stars and gas are distributed at the galactic center compactly, resulting in smaller $r_{\rm e}$.
Meanwhile, $r_{\rm e}$ reaches the upper limit ($0.1r_{\rm vir}$) when the main galaxies interact with satellite galaxies.
In addition, $r_{\rm e}$ gradually increases with the virial radius ($\propto \Mh^{1/3}(1+z)^{-1}$) as redshift decreases.
Finally, the galactic size becomes $\sim 1~{\rm kpc}$ at $z\sim 6$, and more massive galaxy has a larger size. 

Understanding the size distribution of high-$z$ galaxies is crucially important for the incompleteness correction in deriving a luminosity function (LF).
Extended galaxies are more unlikely to be detected for a given magnitude limit, or some fraction of their flux is lost due to the limited sensitivity.
This affects the estimation of faint-end slope of UV LFs \citep{Grazian11}.

\begin{figure}
\begin{center}
\includegraphics[width=\columnwidth]{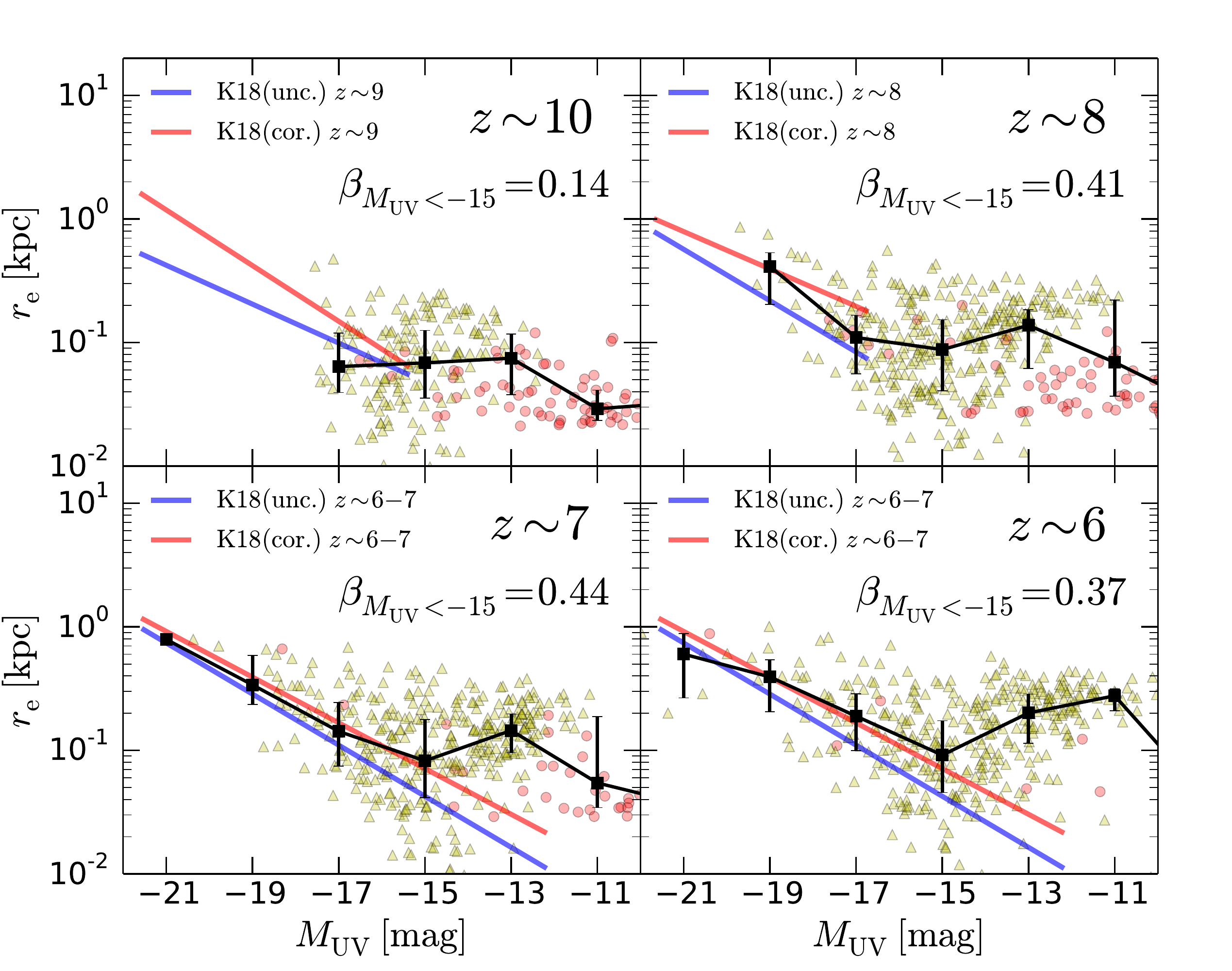}
\caption{
Size--luminosity relation of all of galaxies within the zoom-in region of Halo-11 (red circles) and Halo-12 (yellow triangles) at $z\sim 6,~7,~8,$ \& $10$.
Black squares and error-bars represent the medians and the quantiles within each magnitude bin of  $\Delta M_{\rm UV}=2.0~{\rm mag}$.
We compute the slope $\beta$ for the relation $r_{\rm e}\propto L^{\beta}$ for the simulated galaxies with brighter than $M_{\rm UV}=-15$.  The red and blue lines show the fit to the observational result by \citet[][]{Kawamata18}, with and without the incompleteness correction, respectively.
}
\label{fig:sizedist}
\end{center}
\end{figure}

Figure~\ref{fig:sizedist} shows the size--luminosity (RL) relation of all satellite galaxies in the zoom-in regions of Halo-11 and Halo-12 at $z\sim 6- 10$.
\cite{Kawamata18} conducted simultaneous maximum-likelihood estimation of LF and  RL relation for galaxies at $z>6$, and they found that the slope $\beta$ for the RL relation $r_{\rm e} \propto L^{\beta}$ is $\sim 0.4$.
For the bright galaxies ($M_{\rm UV}<-15$) at $z\sim 6-8$,  our results match the observational data well.

On the other hand, the faint galaxies ($M_{\rm UV}\gtrsim -15$) have a larger dispersion and their sizes are away from the observational fit. 
This can be related to the intermittent star formation histories of low-mass galaxies.
Stellar feedback induces the angular momentum redistribution of gas disc and make the stellar distribution extended \citep[e.g.,][]{El-badry16}.
Another reason is the rapid change of local gravitational potential due to feedback. If star clusters are virialized with the local gravitational potential, they can spread out with the gas outflow. 
In addition, the quenching time of star formation becomes longer as the galaxy mass decreases, resulting in larger size decided by extended residual stars (in contrast, when galaxies have star formation, UV light is dominated by the star-forming region, resulting in smaller $r_{\rm e}$).
These effects might turn the RL relation up at the faint end ($M_{\rm UV} \gtrsim -13$).
Thus, we suggest that some fraction of low-mass galaxies could be lost in observation due to the extended stellar distribution and faint surface brightness.   This could change the faint-end slope of observed LFs. 
To avoid picking up multiple clumpy regions of low-mass galaxies which could boost up the value of $r_{\rm e}$, we did not include the galaxies that are larger than 10 per cent of virial radii. 

Note that, however, the galactic size or `compactness' can be affected by star formation model as described in Y17.
In addition, \cite{Wyithe11} pointed out that the RL relation is also affected by SN feedback model.
We also present the size evolution of Halo-11-lowSF to show the difference from that of Halo-11.
The inefficient star formation induces to form very dense and compact gas clumps at galactic centers, which efficiently traps UV photons.
Thus $r_{\rm e}$ stays at lower value for most of the time. 
We discuss more details of model dependence in Section~\ref{sec:model}.

%
%
\section{Discussion}
\label{sec:discussion}

%
%
\subsection{Dependence on star formation and feedback models}
\label{sec:model}

We have studied the radiative properties of different haloes simulated with the same star formation and feedback models.
Here we investigate how the radiative properties change 
if we decrease the star formation efficiency (Halo-11-lowSF) or turn off the SN feedback (Halo-11-noSN).
Recent observations suggested that the amplitude factor of Kennicutt-Schmidt law was much higher than the local galaxies for merging galaxies \citep{Genzel10} or high-redshift galaxies \citep{Tacconi13}.
Therefore Y17 used a high amplitude factor of $A=1.5 \times 10^{-3}~\Msun \; {\rm yr^{-1} \; kpc^{-1}}$ which was higher than that of the local galaxies by a factor 10 \citep[see also][]{Khochfar11,Silk13}.  
Halo-11-lowSF uses the amplitude factor same as the local galaxies, i.e., $A=1.5 \times 10^{-4}~\Msun \; {\rm yr^{-1} \; kpc^{-1}}$. 

\begin{figure}
\begin{center}
\includegraphics[width=\columnwidth]{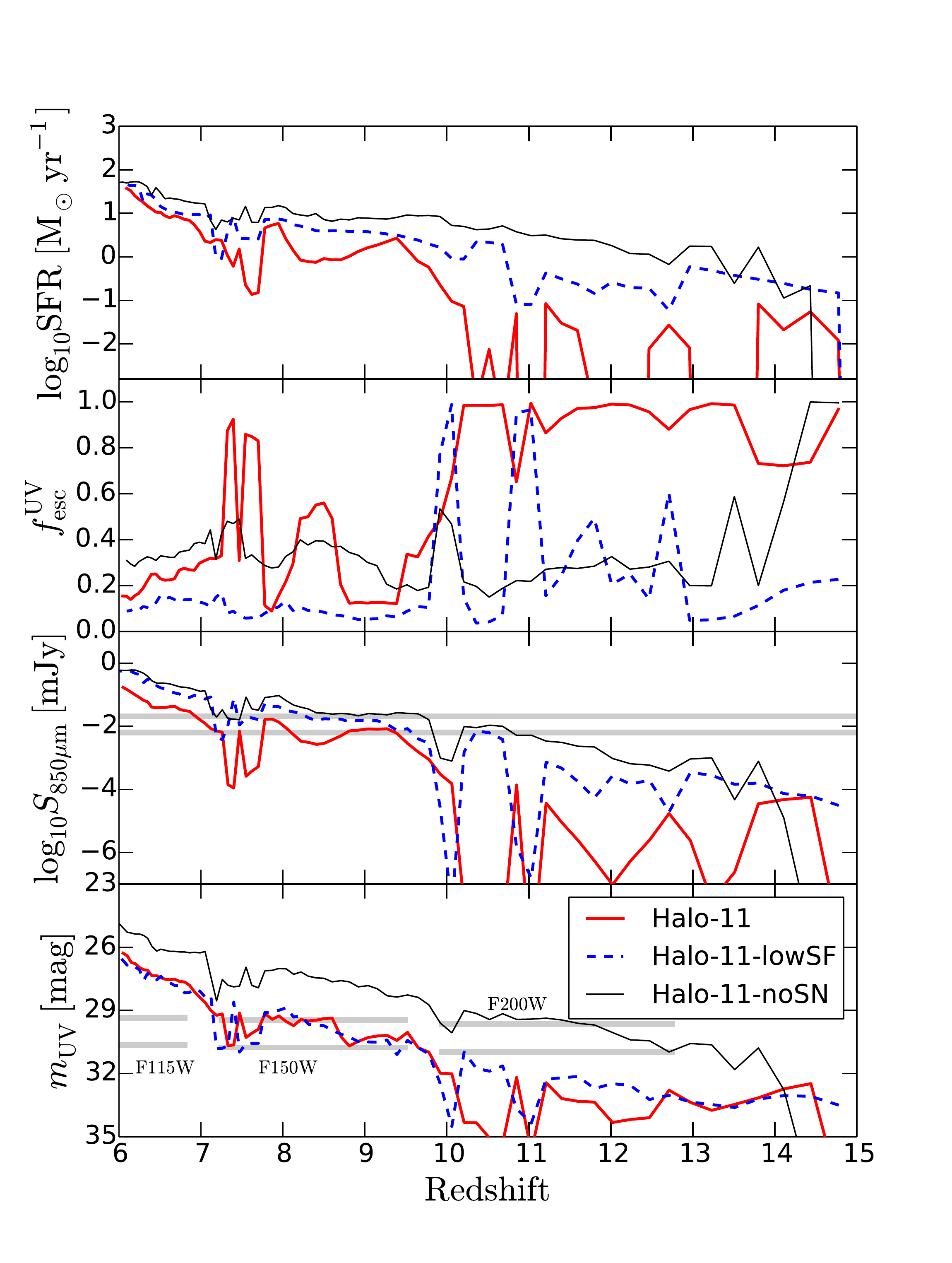}
\caption{
Same as Fig.~\ref{fig:Mhalo} but for the cases of Halo-11, Halo-11-lowSF and Halo-11-noSN.
}
\label{fig:models}
\end{center}
\end{figure}

Figure~\ref{fig:models} shows the star formation histories and radiative properties of Halo-11, Halo-11-lowSF, and Halo-11-noSN. 
In the case of Halo-11-lowSF, the star formation occurs continuously because the central gas density is very high, 
which makes the SN feedback inefficient due to the efficient radiative cooling (see the details in Y17).  Therefore, the gas clouds are not completely disrupted and the continuous star formation is allowed.  
The star-forming gas clouds also efficiently absorb stellar UV radiation, resulting in the low escape fraction as shown in the second panel of the figure. 
In the case of Halo-11-noSN, most of the gas is rapidly converted into stars, resulting in the lower gas density than that of Halo-11-lowSF. 
Therefore $f_{\rm esc}^{\rm UV}$ of Halo-11-noSN is higher than that of Halo-11-lowSF.
Also $f_{\rm esc}^{\rm UV}$ does not change with time significantly due to lack of feedback. 

In Halo-11-lowSF and Halo-11-noSN,  the SFR is higher than Halo-11, and $f_{\rm esc}^{\rm UV}$ is lower, resulting in higher sub-mm flux than Halo-11. 
In addition, the difference between Halo-11-lowSF and Halo-11-noSN clearly appears in the UV magnitude due to different escape fraction.
Thus we argue that the radiative properties of first galaxies sensitively depend on the star formation and feedback models.

%
%
\section{Summary}
\label{sec:summary}

In this paper we have studied the  radiative properties of first galaxies at $z=6-15$ by combining cosmological hydrodynamic simulations and radiative transfer calculations.
Using zoom-in initial conditions, we follow the formation and evolution of three haloes: Halo-10 ($\Mh = 2.4 \times 10^{10}~\Msun$), Halo-11 ($\Mh = 1.6 \times 10^{11}~\Msun$), and Halo-12 ($\Mh = 0.7 \times 10^{12}~\Msun$) at $z=6$. Our major findings are as follows:

\begin{enumerate}
\item 
In the first galaxies, the SN feedback ejects most gas and dust from galaxies, resulting in the intermittent star formation history. 
This causes the large fluctuation of escape fraction of UV photons. 
The escape fraction of Halo-11 changes in the range of $\sim 0.2 - 0.8$ at $z<10$.
As the halo becomes more massive, the fluctuation is suppressed, and the escape fraction remains low at $\lesssim 0.2$. 
The transition redshifts are $\sim 8.5$ for Halo-12 and $\sim 7.5$ for Halo-11. 
In the case of Halo-10, the escape fraction keeps fluctuating down to $z=6$.

\item 
Stellar UV radiation absorbed by dust is re-processed into IR thermal emission. 
Therefore, the IR flux from galaxies also change with time as the UV escape fraction fluctuates.
This fluctuation of the IR flux affects the detectability in sub-mm observations.
Using all satellite galaxies within the zoom-in regions, we calculate the detectability by ALMA telescope.
If we set the detection threshold to $0.1\,\rm mJy$ at $850\,\rm \mu m$, the detectability are $\gtrsim 0.5$ for galaxies with the halo mass of $\gtrsim 10^{11}\,\Msun$ at $z\lesssim 7$. 

\item
We calculate the three-dimensional structure of dust temperature, and derive SEDs. 
By using the peak wavelength of infrared flux, we estimate the typical dust temperature of modeled galaxies. 
The galaxies with $L_{\rm IR} \lesssim 10^{11}\,\Lsun$ have $T_{\rm d}\sim 60\,\rm K$ that is higher than that of observed galaxies at $z < 3$ \citep{Hwang10}. 
Since it is difficult to measure the dust temperature of observed galaxies at $z \ge 6$, 
the dust temperature of $\sim 40\,\rm K$ is frequently assumed in the observations \citep[e.g.][]{Watson15}. 
Our simulation suggests that the dust temperatures for high-$z$ galaxies are somewhat higher than the assumed ones, which will change the estimated dust masses and SFRs.

\item The half-light radius ($r_{\rm e}$) at UV wavelength fluctuates in the range of $r_{\rm e}/r_{\rm vir} \sim 0.02-0.1$, and it increases with time.
At $z\sim 6-8$, $r_{\rm e}$ becomes $\sim 1\,{\rm kpc~(physical)}$,
and the bright galaxies in our simulations ($M_{\rm UV}<-15$) has the relation $r_{\rm e}\propto L^{\beta}$, where $\beta\sim 0.4$. These are consistent with the observations of \citet{Kawamata18}.
When the surface brightness of galaxies is extended, some parts of UV flux can be lost below observational threshold due to cosmological surface brightness dimming.  

\item We compare our fiducial model (Halo-11) with the different models with a low star formation amplitude factor (Halo-11-lowSF) and without the SN feedback (Halo-11-noSN).
In cases of Halo-11-lowSF and Halo-11-noSN, massive dusty gas accumulates at the galactic center due to the weak or no SN feedback. 
This causes higher star formation rate and lower UV escape fraction than the fiducial model.  
Therefore, the IR flux of Halo-11-lowSF and Halo-11-noSN is higher than that of Halo-11. Thus we argue that the star formation and feedback models for the first galaxies could be constrained by future observations. 

\end{enumerate}

In this work, we have investigated the time evolution of radiative properties of the first galaxies, and suggested that the first galaxies rapidly changed from UV-bright to IR-bright due to intermittent star formation and SN feedback. However, the sample of simulated galaxies is not large enough to make a statistical comparison to observations. 
We will obtain a larger galaxy sample using larger-scale and higher-resolution cosmological simulations, and present the statistical properties, e.g. the luminosity function and cosmic star formation rate density, taking the detailed radiative processes into account in our future paper.
Furthermore, the radiative properties of galaxies depend on the dust properties, e.g. size distribution and compositions \citep{Aoyama17,McKinnon18}, therefore we will implement a model of dust destruction and production in cosmological simulations, coupled with radiative transfer calculations. 

%
%
%
%
\section*{Acknowledgments}
Numerical computations were carried out on the Cray XC30 \& XC50 at the Center for Computational Astrophysics, National Astronomical Observatory of Japan, and the {\it Octopus} at the Cybermedia Center, Osaka University. 
This work is supported in part by the MEXT/JSPS KAKENHI Grant Number JP17H04827 (H.Y.), 18H04570 (H.Y.) and JP17H01111 (K.N.).
KN thanks to Dr. Kawamata for useful discussions on galaxy size evolution.
%
%
\bibliographystyle{mn}

\bibliography{refs}


%
%
\appendix


\label{lastpage}

\end{document}